\documentclass{aa}  
\usepackage{ifpdf}
\usepackage{graphicx}
\usepackage{epstopdf}
\usepackage{txfonts}
\usepackage{natbib}
\begin{document}
   \title{Deuterium chemistry in the Orion Bar PDR\thanks{Based on observations with the IRAM 30\,m telescope at Pico Veleta (Spain) and the Atacama Pathfinder EXperiment (APEX) telescope. IRAM is funded by the INSU/CNRS (France), the MPG (Germany) and the IGN (Spain). APEX is a collaboration between the Max-Planck-Institut f\"ur Radioastronomie, the European Southern Observatory, and the Onsala Space Observatory. }}

   \subtitle{``warm" chemistry starring CH$_2$D$^+$}

   \author{B. Parise
          \inst{1}%\fnmsep%\thanks{Humboldt fellow}
          \and
          S. Leurini\inst{1,2}
          \and
          P. Schilke\inst{1}
          \and
          E. Roueff\inst{3}
          \and
          S. Thorwirth\inst{1}
          \and
          D. C. Lis\inst{4}
                 }

   \offprints{B. Parise}

   \institute{Max Planck Institut f\"ur Radioastronomie, Auf dem H\"ugel 69, 53121 Bonn, Germany\\
              \email{bparise@mpifr-bonn.mpg.de}
         \and European Southern Observatory, Karl-Schwarzschild-Str. 2, 85748 Garching bei M\"unchen, Germany 
         \and LUTH, Observatoire Paris-Meudon, 5 Place Jules Janssen, 92195 Meudon Cedex, 
France
         \and  
California Institute of Technology, MC 301-17, Pasadena, CA 91125, USA                   }

   \date{Received xxx; accepted xxx}

% \abstract{}{}{}{}{} 
% 5 {} token are mandatory
 
  \abstract
   {High levels of deuterium fractionation in gas-phase molecules are usually associated with cold regions, such as prestellar cores. Significant fractionation ratios are also observed in hot environments such as hot cores or hot corinos, where they are believed to be produced by the evaporation of the icy mantles surrounding dust grains, and thus are remnants of a previous cold (either gas-phase or grain surface) chemistry.    
   The recent detection of DCN towards the Orion Bar, in a clump at a characteristic temperature of 70\,K, has shown that high deuterium fractionation can also be detected in PDRs. The Orion Bar clumps thus appear as a good environment for the observational study  of deuterium fractionation in luke-warm gas, allowing to validate chemistry models in a different temperature range, where dominating fractionation processes are predicted to be different than in cold gas ($<$\,20\,K). }
   {We aimed at studying observationally in detail the chemistry at work in the Orion Bar PDR, to understand if DCN is produced by ice mantle evaporation, or is the result of warm gas-phase chemistry, involving the CH$_2$D$^+$ precursor ion (which survives higher temperatures than the usual H$_2$D$^+$ precursor).  }
   {Using the APEX and the IRAM 30\,m telescopes, we targetted selected deuterated species towards two clumps in the Orion Bar.  }
   {We  confirmed the detection of DCN and detected two new deuterated molecules (DCO$^+$ and HDCO) towards one clump in the Orion Bar PDR. Significant deuterium fractionations are found for HCN and H$_2$CO, but a low fractionation in HCO$^+$. We also give upper limits for other molecules relevant for the deuterium chemistry.  }
   {We argue that grain evaporation in the clumps is rather unlikely to be dominant, and we find that the observed deuterium fractionation ratios are consistent with predictions of pure gas-phase chemistry models at warm temperatures (T $\sim$ 50\,K). We show evidence that warm deuterium chemistry driven by CH$_2$D$^+$ is at work in the clumps.  }


   \maketitle
%
%________________________________________________________________

\section{Introduction}

Despite the low deuterium abundance in the universe \citep[D/H $\sim$ 10$^{-5}$,][]{Linsky03}, high abundances of deuterated molecules have been observed in dark clouds and star forming regions in the last few years, with detections of molecules containing up to three atoms of deuterium (ND$_3$: \citealt{Lis02,vanderTak02}, and CD$_3$OH: \citealt{Parise04}), with noteworthy fractionation effects \citep[CD$_3$OH/CH$_3$OH\,$\sim$\,1\%,][]{Parise04}.

Formation of such highly-deuterated molecules requires specific physical conditions, which makes them very interesting probes of the environments where they are observed. In molecular clouds, deuterium is mainly locked into molecular HD. Efficient transfer of deuterium from this reservoir at the low temperature of dark clouds has to occur through ion-molecule reactions, and it has long been known that only a few ions react fast enough with HD to compete against electron recombination: H$_3^+$, CH$_3^+$ \citep{Huntress77} and C$_2$H$_2^+$ \citep{Herbst87}. The deuterated isotopologues of these three ions are thus considered to be the precursors of deuterium fractionation in the gas phase. The transfer of deuterium from the HD reservoir to other molecules is initiated through the following exothermic reactions: 
$$ {\rm H_3^+ + HD \rightarrow H_2D^+ + H_2  ~~~~~~~~~~~~(1)}$$
\vspace{-0.5cm}
$$ {\rm CH_3^+ + HD \rightarrow CH_2D^+ + H_2   ~~~~~~~(2)}$$
\vspace{-0.5cm}
$$ {\rm C_2H_2^+ + HD \rightarrow C_2HD^+ + H_2   ~~~~~\,\,(3).}$$
H$_2$D$^+$, CH$_2$D$^+$ and C$_2$HD$^+$ then transfer their deuterium to the other species through ion-molecule reactions.
Exothermicities are  232\,K \citep{Gerlich02} for reaction (1), $\sim$\,390\,K \citep{Asvany04} for reaction (2) and $\sim$\,550\,K \citep{Herbst87} for reaction (3), so that the reverse reactions are inhibited at low temperatures. Efficient transfer of deuterium to molecules has been widely observed in cold regions where high levels of CO depletion are present, such as dark clouds or prestellar cores. In these environments, H$_3^+$ is predominantly responsible for the fractionation. Reaction (2) is thought to be predominant at slightly higher temperature (T$\sim$ 30-50\,K), when (1) is not efficient anymore due to an increased importance of its reverse reaction. Although the chemistry involving H$_2$D$^+$ is now basically understood, thanks to the numerous detections of H$_2$D$^+$ in prestellar cores \citep[e.g.][]{Caselli03,Caselli08} and the parallel development of chemical models \citep[e.g. ][]{Roberts03, Walmsley04, Flower04, Pagani09}, the contribution of the CH$_2$D$^+$ chemistry has so far not been {\bf {\it observationally}} investigated, due to the lack of observations targetting intermediate temperature sources, warm enough so that the CH$_2$D$^+$ contribution becomes significant 
relative to H$_2$D$^+$, and cold enough for the chemistry not to be dominated by ice evaporation.

Recently, during an unbiased spectral survey of the Orion Bar using the APEX telescope, we detected a deuterated molecule (DCN) for the first time in a molecular clump in a Photon-Dominated Region \citep[hereafter PDR, ][]{Leurini06b}. This was however unexpected due to the high temperature (T\,$\sim$\,70\,K) characteristic of this clump. The fractionation ratio deduced from these observations is 0.7\,\%, a value intermediate between the one observed in dark clouds \citep[L134, 5\%, ][]{Turner01} and hot cores (Orion, 0.1\%, \citealt{Schilke92}; a sample of hot cores, 0.1-0.4\%, \citealt{Hatchell98}).

DCO$^+$ was not detected in this survey, with an upper limit on the DCO$^+$/HCO$^+$ ratio of $\sim$ 0.1\,\% (see below), although observations in other environments point to similar fractionation ratios for the two species (dark cloud L134: Tin\'e et al. 2000, Turner 2001; low-mass protostar IRAS16293: van Dishoeck et al 1995). We proposed that this may be an indication that chemistry involving CH$_2$D$^+$ as precursor for deuterium transfer to molecules is at work in the Orion Bar, making it a reference environment to study in further detail reactions involving these routes. This possibility was confirmed by the theoretical modelling study of \citet{Roueff07}.
Since the DCN detection by \citet{Leurini06b}, DCO$^+$ was detected towards the Horsehead PDR \citep{Pety07}, in a cold (10-20\,K) condensation shielded from the UV illumination. The observations we present in this paper are targetting warmer regions than the condensation observed by \citet{Pety07}.  

We present here a detailed investigation on the deuterium chemistry at work in the dense clumps in the Orion Bar, based on observations with the APEX and IRAM 30m telescopes. Observations are described in section 2, the physical conditions in the clumps (temperature and H$_2$ densities) are derived in section 3, the relative abundances and fractionations for  targetted molecules are determined in section 4, and the chemistry is discussed in section 5.

%__________________________________________________________________

\section{Observations}

The Orion Bar, located in the Orion A molecular cloud, is a well studied PDR, mostly because of its nearly edge-on morphology. Its distance is estimated to 414 pc \citep{Menten07}. It is illuminated by the young trapezium stars, located some 2$'$ to the north-west. Previous studies have shown that this PDR has a clumpy structure (see section \ref{clumpy}). The interferometric observations of \citet{Lis03} displayed a series of molecular clumps, as traced by the H$^{13}$CN(1--0) transition, located in the cloud some 10$''$ behind the H/H$_2$ transition. In the present  study we use the nomenclature defined by \citet{Lis03} to refer to the clumps. 

We present in this section the APEX and IRAM 30m observations of the Orion Bar, targetting particularly clump 1 and 3 of \citet{Lis03}. Unless otherwise stated, the transition frequencies of the molecules are taken from the Cologne Database for Molecular Spectroscopy \citep{Muller01, Muller05}. Table \ref{IRAMdata} summarizes the observations presented in this paper.

\subsection{APEX observations}

Using the APEX telescope on Chajnantor (Chile), we mapped the Orion Bar in the DCN(4--3) and H$^{13}$CN(4--3) transitions.  The double-sideband APEX2a receiver \citep{Risacher06} was tuned to 289.0000\,GHz (DCN) 
and 345.3397\,GHz (H$^{13}$CN), and connected to the two units of the FFTS backend \citep{Klein06}, each with 8192 channels, leading to a velocity resolution of 0.13 and 0.11 km/s respectively over the two times 1\,GHz bandwidths. The APEX beamsize is 21$''$ (respectively 18$''$) at 289\,GHz (respectively 345\,GHz).

The (0$''$,\,0$''$) position of the map is $\alpha$(2000)=05$^h$35$^m$25.3$^s$, $\delta$(2000)=$-$05$^\circ$24$'$34.0$''$, corresponding to the "Orion Bar (HCN)" position of \citet{Schilke01}, the most massive clump seen in H$^{13}$CN \citep{Lis03}, as well as the target of the spectral survey of \citet{Leurini06b}.  
The maps were obtained using the on-the-fly mode, with a dump every 6$''$. The reference position was taken at the (600$''$,\,0$''$) offset position from the center of the map. 

The observations were performed between July 19$^{\rm th}$ and August 2$^{\rm nd}$, 2006, under very good to good  weather conditions (with a precipitable water vapor ranging from 0.3 to 1.5 mm). The typical DSB system temperatures were 115 and 200 K at 289 and 345\,GHz respectively.

Several CH$_3$OH(6-5) transitions were also present in the DCN(4--3) setup. DCN(5--4) and HNC(4--3) were observed towards the two clumps on June 28$^{\rm th}$, 2007, with T$_{\rm sys}$ around 200\,K. 

Observed intensities were converted to T$_{\rm mb}$ using ${\rm T_{mb}= T^*_a / \eta_{\rm mb}}$ where $\eta_{\rm mb}$ = 0.73 \citep{Guesten06}.

We focus here on the observations towards clumps 1 and 3, and analyze the spatial distribution across the Bar in a later paper (Parise et al. in prep).

\subsection{IRAM 30m observations}

We targetted the Orion Bar during three observing runs at the IRAM 30\,m telescope. During the first run we made use of the ABCD receivers, targetting only the two clumps in the Bar. The second run consisted of mapping the Bar with the HERA receiver. Some complementary data on the two clumps were also acquired as part of a third run. 

\subsubsection{Single pixel receivers} 

Using the IRAM 30\,m telescope, we observed different species toward the two brightest H$^{13}$CN clumps of the Orion Bar ---``clump 1" at offset position (0$''$,\,0$''$) and ``clump 3" at position ($-$50$''$,\,$-$40$''$), as denoted by \citet{Lis03}. Besides observing different transitions of DCN to constrain the excitation, and looking for a lower excitation line of DCO$^+$ than the one that was not detected with APEX, we selected for our search molecules that can be synthesized in the gas phase via channels involving the CH$_2$D$^+$ ion. Such molecules include HDCO and C$_2$D \citep{Turner01}. We also searched for CH$_2$DOH and HDO, in order to constrain any possible ice chemistry contribution.

The observations were performed from September 29th to October 7$^{\rm th}$, 2006, under variable weather conditions. Four receivers were used simultaneously to observe two different frequency bands (either in the AB or CD setup). The observed lines are listed in Table \ref{IRAMdata}. The receivers were connected to the VESPA correlator in parallel mode, leading to different velocity resolutions depending on the transition. Additionally, the Bonn Fourier Transform Spectrometer \citep{Klein06} was connected to two of the receivers, providing a 850\,MHz bandwidth. 

Some additional data (DCN(3--2) as well as integration on CH$_2$DOH towards both clumps) were taken during a third observing run in May 2008, under poor (CH$_2$DOH data) to moderate (DCN data) weather conditions. 

\subsubsection{HERA observations}

The Orion Bar was mapped in selected methanol and formaldehyde transitions using the HERA receiver, a heterodyne array consisting of two arrays of 3$\times$3 pixels with 24$''$ spacing. The observations were performed during the winter 2007 HERA pool observing period. The full dataset is presented in \citet{Leurini09}. Here we analyse the methanol observations towards the two clumps to derive their physical properties.

\bigskip

All intensities of observations from the IRAM 30\,m telescope were converted to T$_{\rm mb}$ using ${\rm T_{mb}= \frac{F_{eff}}{B_{eff}}  T^*_a }$ where ${\rm B_{eff}}$  is the main beam efficiency, and ${\rm F_{eff}}$ is the forward efficiency. The main beam efficiencies decrease from 78\% to 50\% between 87 and 241\,GHz\footnote{See http://www.iram.es/IRAMES/telescope/telescopeSummary/ telescope\_summary.html}. Forward efficiency is 95\% at 3\,mm, 93\% at 2\,mm and 91\% at 1.3\,mm.

\begin{table}[!h]
\caption{Summary of the observations.}
\begin{tabular}{@{\extracolsep{-5pt}}lllcc}
\noalign{\smallskip}
\hline
\hline
\noalign{\smallskip}
Transition & Frequency &  Telescope  &  Beamsize   &  Targetted\\
                   &   (GHz)       &                        &                       &    Clump         \\
\noalign{\smallskip}
\hline
\noalign{\smallskip}
DCN(2--1) & 144.8280015  &  IRAM 30\,m   &  17$''$  &  1, 3  \\
DCN(3--2)  & 217.2385378  &  IRAM 30\,m  & 11$''$ &  1, 3  \\
DCN(4--3)  & 289.6449170 & APEX & 21$''$ &  map  \\
DCN(5--4)  & 362.0457535 & APEX & 17$''$  &  1, 3  \\
DCO$^+$(2--1)  & 144.0772890  &  IRAM 30\,m  & 17$''$  & 1, 3 \\
HDCO(2$_{11}$-1$_{10}$) & 134.2848300 &  IRAM 30\,m  & 18$''$  & 3 \\
CH$_2$DOH(2--1) & 89.3 (band) & IRAM 30\,m & 28$''$ & 1, 3\\
HDO(2$_{1,1}$-2$_{1,2}$) & 241.5615500$^1$  &  IRAM 30\,m & 10$''$ & 3\\
C$_2$D(2--1) & 144.3 (band) &  IRAM 30\,m & 17$''$ & 3\\
DNC(2--1) & 152.609774 & IRAM 30\,m & 16$''$ & 3 \\
\noalign{\smallskip}
\hline
\noalign{\smallskip}
H$^{13}$CN(1--0) & 86.3399215 &  IRAM 30\,m  &  28$''$& 1, 3 \\
H$^{13}$CN(3--2) & 259.0117978  & IRAM 30\,m &   9.5$''$ & 1, 3  \\
H$^{13}$CN(4--3) & 345.3397694  & APEX &  18$''$ &    map  \\
H$^{13}$CO$^+$(1--0) &  86.7542884 &  IRAM 30\,m & 28$''$ & 1, 3\\
HCO$^+$(1--0) & 89.1884957  & IRAM 30\,m & 28$''$ & 1, 3 \\
H$_2^{13}$CO(2$_{11}$-1$_{10}$) & 146.6356717 & IRAM 30\,m & 17$''$  & 3 \\
C$^{17}$O(1--0) & 112.36 & IRAM 30\,m & 22$''$ &  3 \\
C$^{17}$O(2--1) & 224.71 & IRAM 30\,m  & 11$''$  &  3 \\
CH$_3$OH(5--4)$^2$ & 241.8 (band) &  IRAM 30\,m & 10$''$ & map \\
CH$_3$OH(6--5) & 290.1 (band)  &  APEX & 21$''$ & 1, 3 \\
CH$_3$OH(1$_{1}$--1$_{0}$ A) & 303.367 & APEX & 21$''$ & 1, 3 \\
HNC(4--3) & 362.63 & APEX & 17$''$ & 1,3 \\  
 \noalign{\smallskip}
 \hline
 \noalign{\smallskip}
\end{tabular}\\
$^1$ JPL database.\\
$^2$ HERA observations (see Leurini et al. (subm) for more detail).\\
\label{IRAMdata}
\end{table}

\section{The clumpy morphology of the Orion Bar} 
\label{clumpy}

The Orion Bar was shown to have an heterogeneous structure, with clumpy molecular cores embedded in an interclump gas. The two-component morphology was first inferred by \citet{Hogerheijde95} and \citet{Jansen95}, because single-dish observations of CS, H$_2$CO and HCO$^+$ could not be fit with a single density component. 
The clumpy structure was later confirmed directly by interferometric observations \citep{YoungOwl00}. The clumps have a density of several 10$^6$ cm$^{-3}$ while the density of the interclump gas is $\sim$\,10$^4$-10$^5$ cm$^{-3}$ \citep{YoungOwl00}. Interferometric maps have shown that H$^{13}$CN is mostly confined to the clumps \citep{Lis03}, which are relatively cold ($\sim$70\,K) compared to the interclump gas (of typical temperature of 150\,K). 

In the following, we discuss the physical parameters of the clumps, based on new observations. We first derive the H$_2$ column density (Section \ref{h2col}), and then the temperature and H$_2$ density based on methanol observations (Section \ref{TnH2}).

\subsection{The H$_2$ column density of the clumps}
\label{h2col}

In this section, we attempt to derive the H$_2$ column density in the two clumps as accurately as possible, in order to be able to compute 
molecular abundances relative to H$_2$ in the following sections. This will then allow us to compare the measured fractional abundances to predictions of chemical models.

\subsubsection{Clump 1}
\label{clump1}

An H$_2$ column density of 9\,$\times$\,10$^{22}$ cm$^{-2}$ averaged on the 18$''$ beam was derived towards clump 1 by \citet{Leurini06b} from analysis of the C$^{17}$O(3-2) emission line, assuming a rotational temperature of 70\,K.

We can also get an estimate of the H$_2$ column density in the clump from the analysis of the dust emission 
observed in the frame of a project targetting clump 1 with the Plateau de Bure interferometer in March, April and December 2004 \citep[follow-up project of the work from][]{Lis03}. The observations were performed in the mosaic mode, with seven fields covering clump 1 in a hexagonal pattern with a central field. The 3\,mm receivers were tuned at the $^{13}$CO(1--0) frequency (110\,GHz), and the 1\,mm receivers targetted H$_2$CO at 218\,GHz. The array configurations, UV coverage and 1\,mm observations are discussed in detail in \citet{Leurini09}. The receiver temperatures at 3\,mm were around 200\,K or better.

Although continuum emission from clump 1 is not detected at 1\,mm, weak emission is detected at 3mm. This suggests that the density profile of the clump is rather smooth, and that its emission is mostly filtered out by the interferometer at 1mm. The integrated intensity measured in a 10$''$ diameter aperture centered on the clump is 0.043 Jy, and 0.12 Jy in a 20$''$ diameter aperture. Assuming T$_{\rm dust}$ = 45\,K (see section \ref{TnH2}), $\beta$\,=\,2 and $\kappa_{\rm 230\,GHz}$\,=\,3.09\,$\times$\,10$^{-1}$\,cm$^2$\,g$^{-1}$ \citep{Ossenkopf94}, we derive N$_{\rm H_2}$\,=\,1.6\,$\times$\,10$^{23}$\,cm$^{-2}$ (resp 1.1\,$\times$\,10$^{23}$\,cm$^{-2}$) averaged in a 10$''$ (resp 20$''$) area centered on the clump.
These values are intermediate and consistent within a factor of 2 both with the column densities derived by \citet{Leurini06b}, and by \citet{Lis03} from H$^{13}$CN observations (2.6\,$\times$\,10$^{23}$\,cm$^{-2}$). In the following, we will use our newly derived value N$_{\rm H_2}$\,=\,1.6\,$\times$\,10$^{23}$\,cm$^{-2}$.

\subsubsection{Clump 3}

Clump 3 was not targetted by the Plateau de Bure observations presented in sec. \ref{clump1}. No continuum was either detected at 3\,mm in the study of \citet{Lis03}, which points to the fact that the density profile of the clump may be again rather smooth. From the H$^{13}$CN observations of \citet{Lis03}, the H$_2$ column density is 1.9\,$\times$\,10$^{23}$\,cm$^{-2}$.

Using the IRAM 30\,m telescope, we targetted the 1--0 and 2--1 transitions of C$^{17}$O towards this clump. The hyperfine structure of the 1--0 transition is clearly resolved. The two observed lines have a rotation temperature of 12\,K, and lead to a N$_{\rm H_2}$ = 8\,$\times$\,10$^{22}$ cm$^{-2}$, averaged on the extent of the clump (8$''$, see below).  

In the following, we therefore assume an intermediate value between the column density as measured from H$^{13}$CN and C$^{17}$O, N$_{\rm H_2}$ = 1.3\,$\times$\,10$^{23}$\,cm$^{-2}$, averaged on the 8$''$ clump.

\begin{figure}[!ht]
\includegraphics[angle=-90,width=8.8cm]{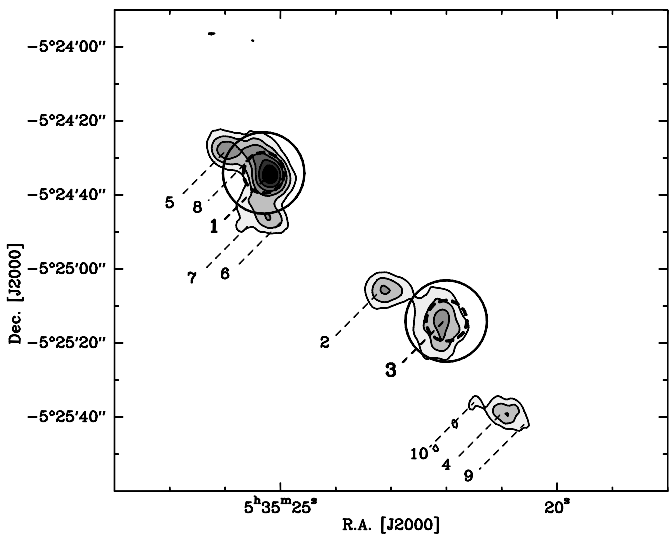}
\caption{In grey scale the H$^{13}$CN (1--0) transition observed with the Plateau
de Bure interferometers by \citet{Lis03}. The numbers indicate the clumps identified by the same authors. 
The solid and
dashed circles outline the beams of the APEX ($\sim\,290$~GHz) and
IRAM~30m ($\sim\,241$~GHz) telescopes respectively.}
\label{beams}
\end{figure}

\subsection{Physical conditions in the clumps : CH$_3$OH analysis}
\label{TnH2}

To determine the properties of the gas in the clumps, we analysed the
methanol emission at 241 and 290~GHz. The beam of the APEX telescope at 290~GHz is almost
twice as large as the one of the IRAM antenna, and therefore samples 
different gas volumes (see Fig.~\ref{beams}). We therefore smoothed the HERA data 
to the resolution of the APEX telescope at 290~GHz. For the analysis, we used the
technique described by \citet{Leurini04} 
for the study of multi-line CH$_3$OH observations, which consists of
modelling all the lines simultaneously with a synthetic spectrum computed
using the Large Velocity Gradient approximation, and comparing it to
the observations. Rest frequencies are from \citet{Xu97}, while the
collisional rates were computed by
\citet{Pottage02,Pottage04}.  The parameters
defining the synthetic spectrum are: source size, kinetic 
temperature, column density, velocity width and velocity offset (from
the systematic velocity of the object).  The line width and the
velocity of the object are not free parameters, but are given as input
values to the model.  Finally, several velocity components, which are
supposed to be non-interacting, can be introduced.

\begin{table}[!ht]
\begin{center}
\caption{Best fit model results from the CH$_3$OH analysis {\bf towards the two clumps}; the uncertainties correspond to the $3\sigma$ confidence level. The methanol column density (N) and fractional abundance relative to H$_2$ (x) is computed averaged on the extent of the clumps.}
\begin{tabular}{lccccc}
\noalign{\smallskip}
\hline\hline
Source  &\multicolumn{1}{c}{S}&\multicolumn{1}{c}{${\rm n_{H_2}}$}&\multicolumn{1}{c}{T}&\multicolumn{1}{c}{N$_{\rm{CH_3OH}}$} &\multicolumn{1}{c}{x$_{\rm{CH_3OH}}$}\\
&\multicolumn{1}{c}{['']}&\multicolumn{1}{c}{[cm$^{-3}$]}&\multicolumn{1}{c}{[K]}&\multicolumn{1}{c}{[cm$^{-2}$]} & \\
\noalign{\smallskip}
\hline
\noalign{\smallskip}
clump 1& 10 & $6^{+4}_{-3}~10^6$&45$^{+47}_{-17}$&$3^{+1}_{-1}~10^{14} $ & 2$\times$10$^{-9}$ \\
\noalign{\smallskip}
clump 3&8&$ 5^{+5}_{-2}~10^6   $&35$^{+17}_{-15}$&$ 3^{+3}_{-1}~10^{14}$ & 2$\times$10$^{-9}$ \\
\noalign{\smallskip}
\hline
\noalign{\smallskip}
\end{tabular}
\label{values}
\end{center}
\end{table}

For our analysis, we modelled the emission towards each clump with a
single component model, and neglected effects due to infrared
pumping.  We used as free parameters the column density of the two
symmetry states of methanol, CH$_3$OH-$A$ and CH$_3$OH-$E$, the
kinetic temperature of the gas, and the H$_2$ density. 
For clump 3, we used a source size of $8''$ as derived from
H$^{13}$CN by \citet{Lis03}.  For clump 1, the source
size derived from the analysis of the H$^{13}$CN emission is $\sim
7''$; however, other clumps partially fall in the $\sim 20''$ beam
size of our observations (see Fig.~\ref{beams}). Therefore, we adopted
a source size of $10''$ which should take into account for the
emission of the other clumps at the edge of the beam. This corresponds
to the assumption that all clumps in the Bar have similar physical
properties.  

The methanol spectra from the two clumps are very similar, although
clump 3 is shifted in velocity of 0.7\,km\,s$^{-1}$ with respect to clump 1
($v_{\rm{ LSR}}$\,$\sim$\,10.0\,km\,s$^{-1}$ at clump 1,
$v_{\rm{LSR}}$\,$\sim$\,10.7\,km\,s$^{-1}$ at clump 3). The spectra are
characterised by narrow lines ($\sim$\,1.2\,km\,s$^{-1}$), and no emission
is detected in transitions with upper level energies corresponding to more than 85~K.  As
already discussed by \citet{Leurini06b}, at the original
velocity resolution of the APEX observations ($\Delta v$\,$\sim$\,0.12\,km\,s$^{-1}$), 
the CH$_3$OH ground state lines toward clump 1
show a double-peaked profile, probably due to the different clumps
sampled by the beam.  Since no double-peaked profile is detected in
the IRAM data ($\Delta v$\,$\sim$\,0.38\,km\,s$^{-1}$) towards the same
position, we smoothed the APEX spectra towards clump 1 to the same
velocity resolution as the IRAM data.  No double-peaked profile is
detected towards clump 3. However, given the low signal to noise ratio
of the APEX observations, we smoothed these data to a resolution of
$0.5$\,km\,s$^{-1}$. The main difference in the methanol spectra of the two
clumps is in the  $1_1 \to 1_0$-$A$ line at 303.4\,GHz which is much stronger
in clump 3. Since this line has a lower level energy of 2.3\,K, this suggests that 
clump 3 is colder than clump~1.

In Table~\ref{values} we report the results of the best fit models 
for both clumps. For the column density, the total value of the column
density for the two symmetry states is given.
The high value for the
spatial density suggests that the gas is close to thermal equilibrium,
as verified by the results of our LVG calculations that show that most lines 
are partially sub-thermal.  We carried out a $\chi^2$ analysis to derive the
uncertainties on the kinetic temperature, density and column density. The values of the $3\sigma$ confidence
levels of each parameters are reported in Table~\ref{values}.

All lines are
well fitted by the model, with the only exception of the $5_1\to
4_1$-$E$ transition. The behaviour of this
line remains unclear:  in massive star forming regions \citep{Leurini07} the
$5_1\to 4_1$-$E$ is observed with an intensity roughly half of that of
the blend of the $5_{\pm 2}\to 4_{\pm 2}$-$E$ lines, as expected since
the three transitions have very similar energies. On the other hand, the $5_1\to 4_1$-$E$ transition in the Orion Bar shows the same 
intensity of the $5_{\pm 2}\to 4_{\pm 2}$-$E$ lines blend, and cannot be fitted by our models. 

For clump 3, the intensities
of the ground state lines in the $6_k \to 5_k$ band, and in the $1_1
\to 1_0$-$A$ line at 303.4~GHz are understestimated. The $1_1 \to
1_0$-$A$ transition is very low in energy, and it is expected to be
more intense in cold regions. This suggests that a two component model, with
a layer of gas at a lower temperature, could be more appropriate
for clump 3. However, the reduced $\chi^2$ for our model is 1.8, which is still reasonable for a
simplified analysis.

\begin{figure}
\includegraphics[width=9cm]{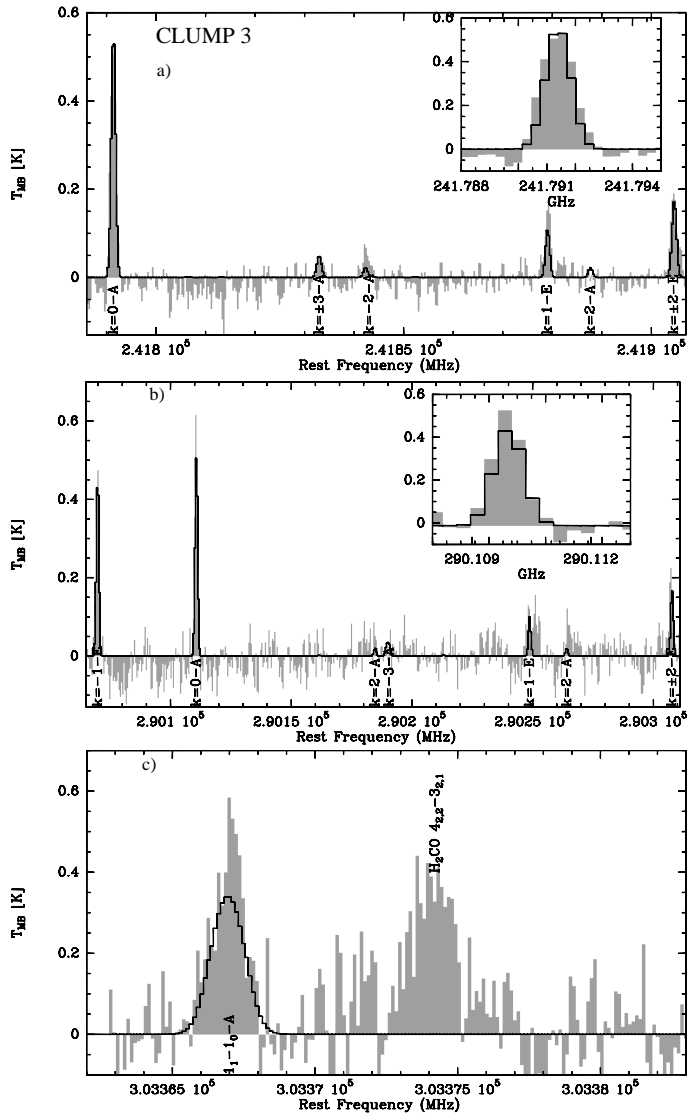}
\caption{Methanol spectra toward clump 3. {\bf a)} 5$\rightarrow$4 band observed with the IRAM 30\,m telescope. {\bf b)  6$\rightarrow$5 band observed with the APEX telescope.  c)  1$_1$--1$_0$-A line (APEX)}. The synthetic spectra resulting from the fit are
overlaid in black. The panel inserts show the $k=0$-$A$ lines in the
two bands to illustrate how well the models fit the observed spectra.}
\label{clump3}
\end{figure}

The spectra for clump 1 and clump 3 are very similar. We therefore present in 
Figure~\ref{clump3} the synthetic spectra for the best fit
model overplotted on the observed spectra for clump 3 only . The second feature in Fig.~2(c) is
identified with (4$_{2,2}\to 3_{2,1}$)~H$_2$CO from the lower side
band.

\begin{table*}[!ht]
\caption{Observational results. Fluxes, FWHM and V$_{\rm lsr}$ are computed through gaussian fitting.}
\begin{tabular}{lcccccc}
\noalign{\smallskip}
\hline
\hline
\noalign{\smallskip}
                    &   \multicolumn{3}{c}{clump 1}   &  \multicolumn{3}{c}{clump 3} \\
Position     &   \multicolumn{3}{c}{(0$''$, 0$''$)}   &  \multicolumn{3}{c}{($-$50$''$, $-$40$''$)} \\                    
\noalign{\smallskip}
\hline
\noalign{\smallskip}
Transition &  $\int$ T$_{\rm mb}$ dv  &  FWHM & V$_{\rm lsr}$ & $\int$ T$_{\rm mb}$ dv & FWHM & V$_{\rm lsr}$ \\
&  {\tiny(K\,km\,s$^{-1}$)}  & {\tiny(km\,s$^{-1}$)} & {\tiny(km\,s$^{-1}$)} &  {\tiny(K\,km\,s$^{-1}$)}  & {\tiny(km\,s$^{-1}$)} & {\tiny(km\,s$^{-1}$)}\\
\noalign{\smallskip}
\hline
\noalign{\smallskip}
DCN(2--1)  &  2.20$\pm$0.04  & 1.9 (hfs) & 10.0$\pm$0.1 & 4.11$\pm$0.05 & 1.3 (hfs) & 10.7$\pm$0.1\\
DCN(3--2)  & 1.62$\pm$0.03 & 1.6$\pm$0.1 & 10.0$\pm$0.1& 1.70$\pm$0.11 & 1.1$\pm$0.2 & 10.6$\pm$0.1\\
DCN(4--3)  &  1.15$\pm$0.07  & 1.3$\pm$0.1 & 10.1$\pm$0.1 &1.75$\pm$0.23 & 1.4$\pm$0.3  & 10.8$\pm$0.1\\
DCN(5--4)  &  0.25$\pm$0.04$^*$ &  1.9$^*$ &  -- & 0.63$\pm$0.10 & 1.7$\pm$0.1 & 10.7$\pm$0.1\\
H$^{13}$CN(1--0) &  1.51$\pm$0.05 &  1.8 (hfs) & 10.1$\pm$0.1 &  1.72$\pm$0.05 & 1.6 (hfs) & 10.6$\pm$0.1 \\
H$^{13}$CN(3--2) &  4.49$\pm$0.04 & 1.1$\pm$0.1 & 10.0$\pm$0.1 & 3.40$\pm$0.04 & 1.3$\pm$0.1 & 10.7$\pm$0.1 \\
H$^{13}$CN(4--3) &  1.66$\pm$0.06 & 1.8$\pm$0.1 & 10.1$\pm$0.1 & 1.72$\pm$0.10 & 2.0$\pm$0.1 &  10.9$\pm$0.1\\
\noalign{\smallskip}
\hline
\noalign{\smallskip}
DCO$^+$(2--1)  &  $<$ 0.03 (3$\sigma$) & 1.8$^\star$ &  --  &   0.12$\pm$0.01 & 1.21$\pm$0.14 & 10.7$\pm$0.1 \\
H$^{13}$CO$^+$(1--0)  & 0.40$\pm$0.02 & 1.75$\pm$0.12 & 10.1$\pm$0.1 & 0.50$\pm$0.02 & 2.20$\pm$0.11 & 10.5$\pm$0.1\\
HCO$^+$(1--0) &  27.8$^{**}$  & 2.08$\pm$0.26 & 9.9$\pm$0.3 & 35.3$^{**}$  &  2.17$\pm$0.03 & 10.5$\pm$0.1 \\
\noalign{\smallskip}
\hline
\noalign{\smallskip}
HDCO(2$_{11}$--1$_{10}$)  & -- & -- & -- &   0.041$\pm$0.005  & 1.20$\pm$0.16 & 10.4$\pm$0.1  \\
H$_2^{13}$CO(2$_{11}$--1$_{10}$) & -- & -- & -- & 0.216$\pm$0.014 & 1.91$\pm$0.15 & 10.8$\pm$0.1 \\
\noalign{\smallskip}
\hline
\noalign{\smallskip}
C$_2$D(2--1)   & --& --& --&  $<$ 0.042 (3$\sigma$) & 1.5$^\star$ &  --  \\ 
CH$_2$DOH(2--1) & $<$ 0.008 (3$\sigma$) & 1.8$^\star$ & -- & $<$ 0.011 (3$\sigma$) & 1.5$^\star$ &   --\\ 
HDO & --& --& --& $<$ 0.093 (3$\sigma$) & 1.5$^\star$ &   -- \\ 
DNC(2--1) & --& --& --& $<$ 0.019 (3$\sigma$) & 1.5$^\star$ &   -- \\ 
HNC(4--3) & -- & -- & -- & 3.03$\pm$0.06 & 2.1$\pm$0.1 & 10.8$\pm$0.1 \\
\noalign{\smallskip}
\hline
\noalign{\smallskip}
\end{tabular}\\
$^*$Line blended with a line coming from the image sideband. Flux was computed with a two-component gaussian fit, keeping the linewidth fixed.\\
$^{**}$Integrated intensity computed without a Gaussian fit (the line is found to be non-Gaussian). \\
$^\star$Assumed width to compute the upper limit on the flux. \\
Uncertainties given on integrated intensities are just the errors given for a gaussian fit, and do not include the calibration uncertainties (assumed to be of the order of 15\%).\\
All 3$\sigma$ upper limits are computed using the following relation: $\int$ T$_{mb}$ dv $<$ 3 $\times$ rms $\times$ $\sqrt{\rm FWHM \times \delta v}$ 
\label{flux}
\end{table*}

\section{Deuterium fractionation in the clumps}

Mapping of the DCN(4--3) emission across the Orion Bar with the APEX telescope shows that DCN emission originates in the H$^{13}$CN clumps imaged by \citet{Lis03}. The overall distribution of DCN in the Bar will be the scope of a forthcoming paper (Parise et al. in prep). Here, we aim at studying the chemistry at work in the clumps, and therefore we target selected transitions of deuterated molecules towards clump 1 and~3.

In the following subsections we study the DCN excitation towards both clumps, present the detection of DCO$^+$ and HDCO towards clump 3, and upper limits on the other molecules.

\subsection{DCN and H${\it ^{13}}$CN excitation}

\begin{figure}[!h]
\includegraphics[width=10cm,angle=-90]{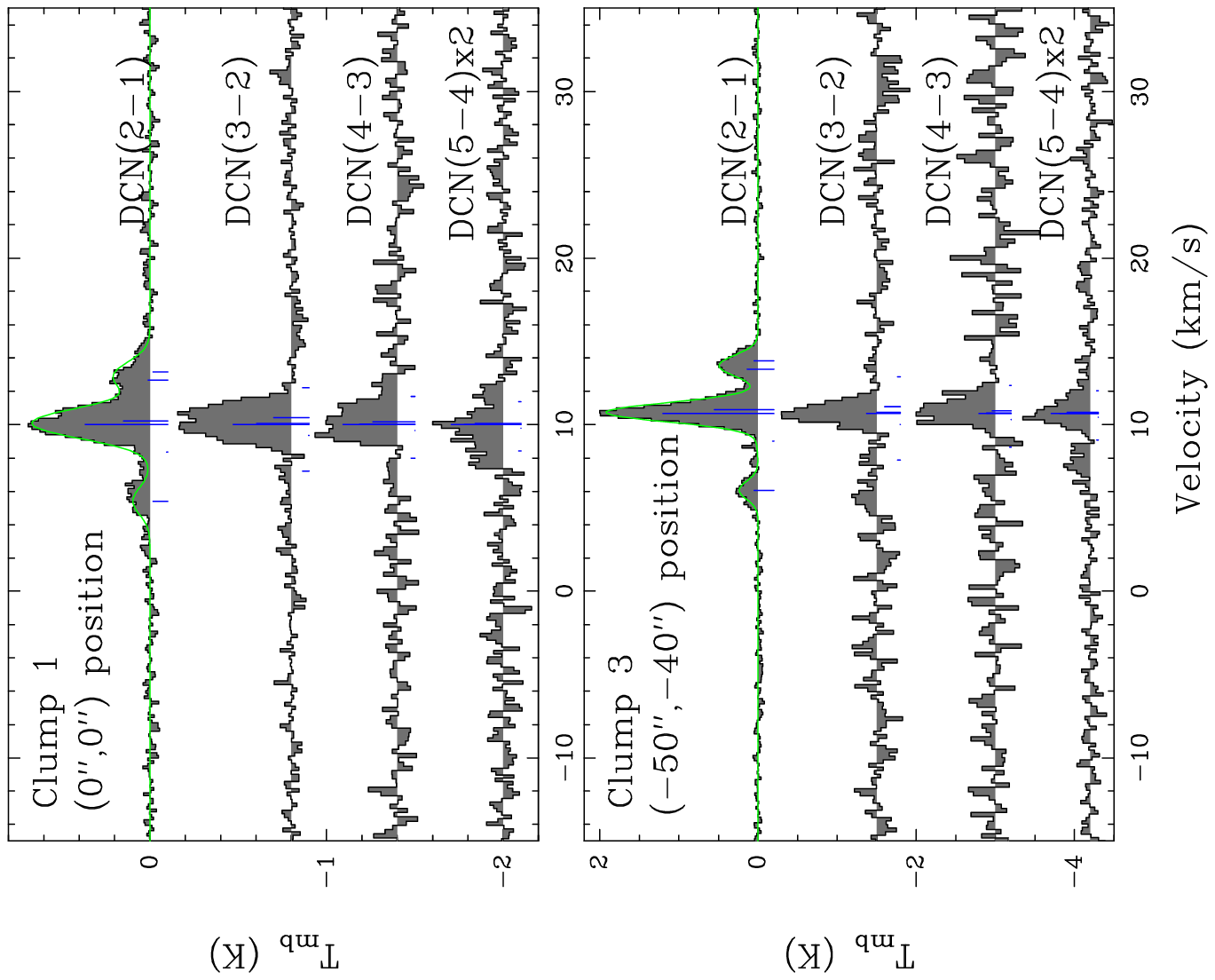}
\caption{DCN spectra observed towards the two groups of clumps. Hyperfine structure fits are displayed in green for the DCN(2--1) lines. A parasite line coming from the image sideband is visible in the DCN(5--4) data (clump 3). Due to the slight velocity difference between the two clumps, the parasite line is blending the DCN(5--4) line on clump 1. }
\label{dcnfig}
\end{figure}

\begin{figure}[!h]
\includegraphics[width=10cm,angle=-90]{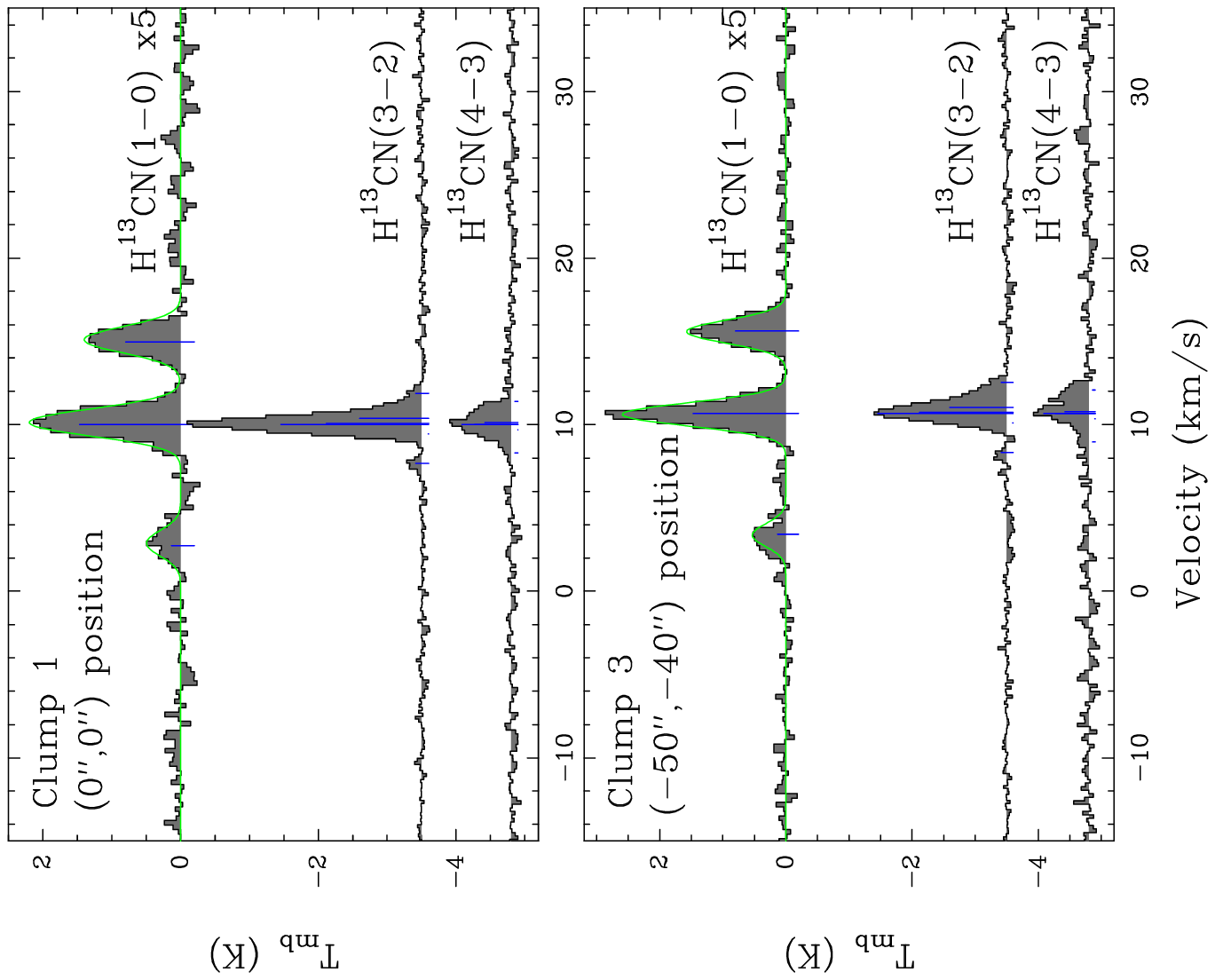}
\caption{H$^{13}$CN spectra observed towards the two groups of clumps. Hyperfine structure fits are displayed in green for the H$^{13}$CN(1--0) lines.}
\label{h13cnfig}
\end{figure}

Figure \ref{dcnfig} (resp. \ref{h13cnfig}) shows the spectra of the several DCN (resp. H$^{13}$CN)  transitions observed towards the clumps with the IRAM 30m 
and the APEX telescopes. In the following subsections we study the excitation of these two molecules, using LTE and LVG methods.

\subsubsection{LTE analysis}

\begin{figure}[!h]
\includegraphics[width=9cm]{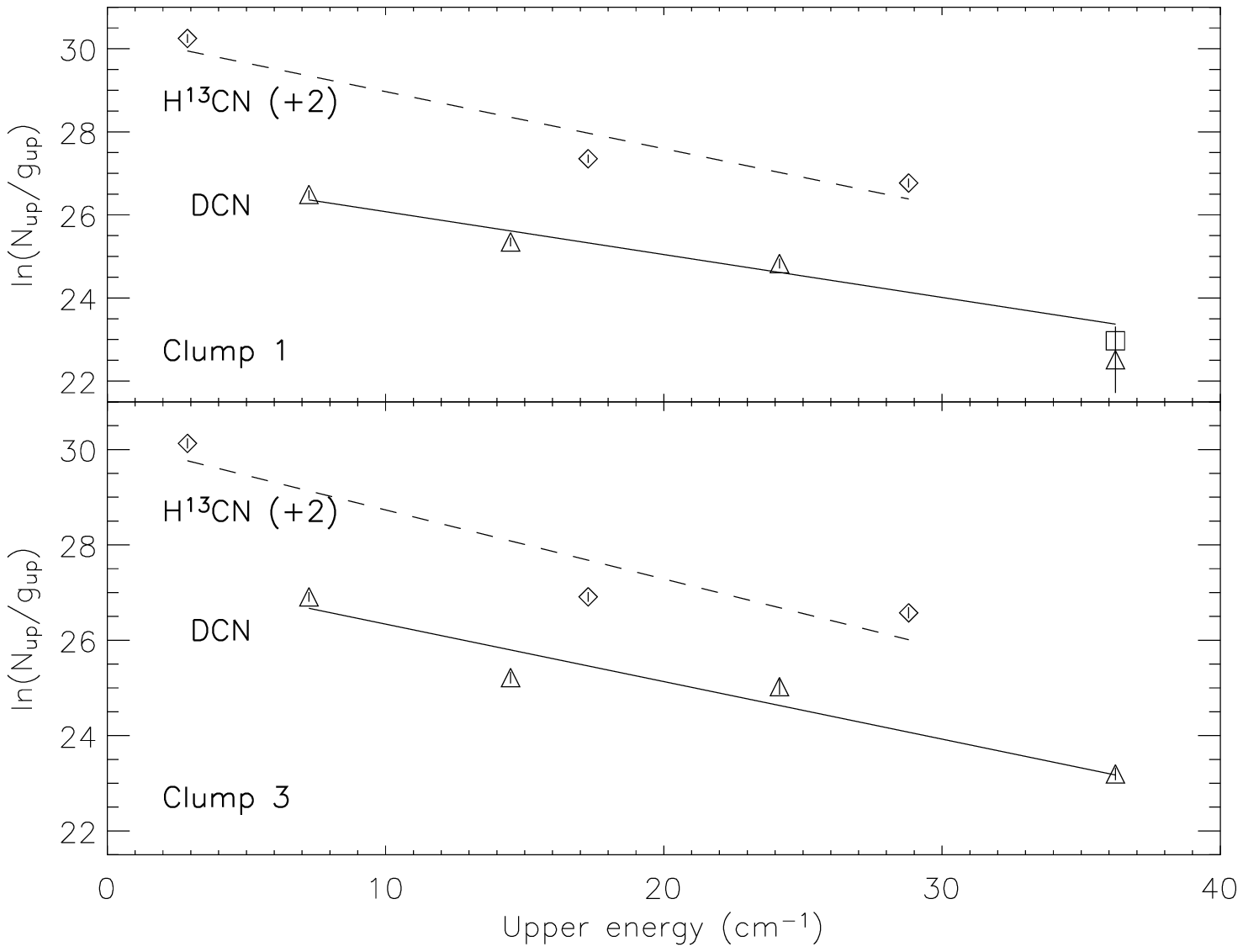}
\caption{Rotational diagrams for DCN (triangles) and H$^{13}$CN (diamonds). The square represents the upper limit for the DCN(5--4) integrated intensity towards clump 1, computed as the total integrated intensity including the parasite line (see text).  }
\label{rotdiag}
\end{figure}

Although the critical densities of the DCN and H$^{13}$CN levels are quite high, several 10$^5$ to several 10$^6$\,cm$^{-3}$, implying that the DCN and H$^{13}$CN level populations may not be in LTE, it is instructive to draw rotational diagrams for the two groups of clumps.  We corrected the observed line intensities for beam dilution effects by assuming a size of 7$''$ for clump 1, and 8$''$ for clump 3 \citep[as derived by][]{Lis03}. The obtained rotational diagrams are presented in Figure \ref{rotdiag}, and the derived rotational temperatures and column densities are listed in Table \ref{rottable}.

\begin{table}
\caption{Results of rotational diagrams and LVG analysis.}
\begin{tabular}{@{\extracolsep{-7pt}}cccccc}
\noalign{\smallskip}
\hline
\hline
\noalign{\smallskip}
\multicolumn{2}{l}{\bf Rotational diagram}\\
Source         & T$_{\rm rot}$({\tiny DCN}) & N$_{\rm DCN}$ & T$_{\rm rot}$({\tiny H$^{13}$CN}) & N$_{\rm H^{13}CN}$  & {\tiny DCN/HCN$^*$} \\
          & {\tiny (K)} & {\tiny (10$^{13}$\,cm$^{-2}$)} & {\tiny (K)} &  {\tiny (10$^{13}$\,cm$^{-2}$)} &  {\tiny (\%)} \\
\noalign{\smallskip}
Clump 1  & 14.0$\pm$1.1  & 1.4$\pm$0.3   & 10.5$\pm$0.4  & 3.1$\pm$0.4   & 0.7$\pm$0.2 \\ 
Clump 3  & 11.9$\pm$0.5  & 1.9$\pm$0.3   & 9.9$\pm$0.4  & 2.5$\pm$0.3   &  1.1$\pm$0.2\\ 
\noalign{\smallskip}
\hline
\noalign{\smallskip}        
\multicolumn{2}{l}{\bf LVG analysis}\\
Source         & T$_{\rm kin}$ {\tiny $[$DCN$]$}& N$_{\rm DCN}$ & T$_{\rm kin}${\tiny $[$H$^{13}$CN$]$} & N$_{\rm H^{13}CN}$  & {\tiny DCN/HCN$^*$} \\
          & {\tiny (K)} & {\tiny (10$^{13}$\,cm$^{-2}$)} & {\tiny (K)} &  {\tiny (10$^{13}$\,cm$^{-2}$)} & {\tiny (\%)} \\
\noalign{\smallskip}
Clump 1 &  18$^{+5}_{-3}$& 1.4$^{+0.4}_{-0.3}$ & 21$^{+3}_{-2}$& 8.2$^{+2.8}_{-2.2}$ & 0.3$\pm$0.1\\
\noalign{\smallskip}   
Clump 3 &  32$^{+13}_{-7}$ & 1.0$^{+0.4}_{-0.3}$  & 26$^{+4}_{-3}$& 1.9$^{+0.4}_{-0.3}$ &  0.8$\pm$0.3 \\
\noalign{\smallskip}
\hline
\noalign{\smallskip}        
\end{tabular}\\
Given error bars are 1$\sigma$. We note that the T$_{\rm kin}$ derived from LVG analysis of DCN and H$^{13}$CN are consistent within the 3$\sigma$ uncertainty range with the T$_{\rm kin}$ values derived from methanol. \\
$^*$Assuming HCN/H$^{13}$CN = 70 \citep{Wilson99}.\\
\label{rottable}
\end{table}

Under the assumption that the lines are optically thin, which is confirmed  by the hyperfine structure analysis (see section \ref{section_hfs}), the deviation from linearity in the rotational diagrams can be due either to the very different beam sizes of the different observations, or to non-LTE effects. The first effect is not completely taken out by correcting for beam dilution, because the source maybe be actually more extended than the smallest beam. This could explain in particular why the DCN(3-2) transition (observed in the smallest beam) is weaker than the other transitions. Although the clumps are very dense ($\sim$\,5\,$\times$\,10$^6$\,cm$^{-3}$), the critical density of the different levels of the molecules is so high that the levels are not populated in LTE. H$^{13}$CN seems even further from LTE than DCN, which is consistent with the fact that critical densities for H$^{13}$CN are around a factor two higher than for DCN. The departure from LTE increases with the increasing upper level energy. The DCN(5--4) transition in clump 1 is in particular subthermally excited compared to the three lower energy transitions. This line is however blended with an unidentified line coming from the image sideband (at 350.554 GHz). The unidentified line is shifted away from the DCN(5--4) lines towards clump 3 due to the slight velocity difference between the two clumps. 

The square in Figure \ref{rotdiag} represents the transition if all flux is considered to come from the DCN(5--4). This point is still too low compared to the three other transitions. It appears that the blending of the line cannot account for the low intensity  of the DCN(5--4) emission, and that the J=5 level population is subthermal. After studying the hyperfine structure of the lower rotational transitions, we will model the line emission using a non-LTE method.

\subsubsection{Hyperfine structure}
\label{section_hfs}

DCN and H$^{13}$CN rotational transitions have an hyperfine structure, caused by the interaction of the electric quadrupole  moment of the N nucleus (I=1) with the molecular field gradient. This causes the transitions to be split into several components, reducing the opacity at the line center. This effect is most important for the lower transitions, DCN(2--1) and H$^{13}$CN(1--0).
From the relative intensity of the hyperfine components, it is possible to derive the opacity of the transition. This was done using the hfs method of the CLASS software (GILDAS package, http://www.iram.fr/IRAMFR/GILDAS/). This method fits the spectrum assuming the same excitation temperature for each hyperfine component. The results of the fits are shown in Table~\ref{hfsresult}. 

\begin{table*}[!ht]
\caption{Results of the hyperfine structure fit.}
\begin{tabular}{llcccccc}
\noalign{\smallskip}
\hline
\hline
\noalign{\smallskip}
Source & Line & v$_{\rm lsr}$ & FWHM & T$_{\rm mb}\times\tau_{\rm tot}$ & $\tau_{\rm tot}$  & T$_{\rm ex} $&  n$_{\rm cr}$ (50\,K) \\
    & & \small{(km s$^{-1}$)} &  {\small (km s$^{-1}$)} &  {\small (K)}   &  &   {\small (K)} &  {\small (cm$^{-3}$)} \\ 
\noalign{\smallskip}
\hline
\noalign{\smallskip}
Clump 1 & DCN(2--1) &  10.01$\pm$0.02 & 1.90$\pm$0.04& 1.2$\pm$0.1 & 0.7$\pm$0.2 & 5.0$\pm$1.8 & 9.1$\times$10$^{5}$ \\
& H$^{13}$CN(1--0) & 10.13$\pm$0.02 & 1.78$\pm$0.07 & 0.92$\pm$0.09& 0.56$\pm$0.41& 4.6$\pm$3.8 & 1.8$\times$10$^{5}$\\
\noalign{\smallskip}
\hline
\noalign{\smallskip}
Clump 3 & DCN(2--1) &  10.66$\pm$0.01 & 1.30$\pm$0.02 & 2.8$\pm$0.1 & 0.1$\pm$0.4 & 31.5$\pm$127 &9.1$\times$10$^{5}$ \\
& H$^{13}$CN(1--0) & 10.62$\pm$0.02 & 1.61$\pm$0.03 & 0.96$\pm$0.02 & 0.10$\pm$0.04 & 12.7$\pm$5.3 &1.8$\times$10$^{5}$\\
\noalign{\smallskip}
\hline
\noalign{\smallskip}
\end{tabular}
\label{hfsresult}
\end{table*}

The total opacity of the DCN(2--1) is found to be low. 
In the case where the opacity is well enough constrained so that the T$_{\rm ex}$ can be accurately derived, the excitation temperature is rather low (Table \ref{hfsresult}). It should be however noted that the derived T$_{\rm ex}$ value is a lower limit because of beam dilution.
The excitation temperature provides a lower limit to the kinetic temperature, as the level populations might not be thermalized, due to the high critical densities of each transition. 
The same holds for H$^{13}$CN(1--0). The transition is optically thin on both positions, and the excitation temperatures are also rather low.

\subsubsection{DCN and H$^{13}$CN column densities}

In order to derive column densities, we used a standard LVG code. 
The hyperfine structure of the lower levels is not explicitely taken into account. However, to account for the reduced opacity due to the hyperfine splitting, we replaced the escape probability by a weighted mean of the escape probabilities of the hyperfine components:  $~~{\rm  \beta = \displaystyle\sum_{i} f_i \frac{1 - e^{-f_i \tau_{tot}}}{f_i \tau_{tot}} }$. 
For both isotopologues, we used collision coefficients calculated for HCN-He by \citet{Wernli09}. These collision coefficients were uniformly scaled by a factor 1.37 (square root of the ratio of the reduced mass of the HCN-He and HCN-H$_2$ systems) to approximate HCN-H$_2$ collisions . This approximation is generally supposed by astronomers to be valid for HCN-pH$_2$ collisions, as para-H$_2$ has a spherical symmetry in its J=0 state. Theoretical computations have shown that the accuracy of this assumption is however limited, the discrepancy for individual rates reaching up to a factor of two for particular systems (H$_2$O--H$_2$/He, \citealt{Phillips96}, CO--H$_2$/He, \citealt{Wernli06}; HC$_3$N--H$_2$/He, \citealt{Wernli07}; SiS--H$_2$/He, \citealt{Lique08}), or even higher (NH$_3$--H$_2$/He, \citealt{Maret09}). This is however the only solution as long as HCN--H$_2$ collision rates are not available.

Input parameters for an LVG code are n$_{\rm H_2}$, T and N$_{\rm mol}$/$\Delta$v. Comparing modelled line temperatures with observations brings another parameter into play, the beam filling factor. On the observational side, we want to fit simultaneously for each molecule the observations of all available transitions and the opacity derived by the hfs method for the lower transition. There are thus 5 observables for DCN and 4 for H$^{13}$CN. 
It is statistically rather meaningless to try to fit a sample of four/five observations with a 4-parameter model. We therefore use independently determined information for the source size \citep{Lis03}, and the H$_2$ density as derived from methanol observations (see section 3.2), and only fit the temperature and the DCN and H$^{13}$CN column densities. 

\begin{figure}[!ht]
\includegraphics[width=8.5cm]{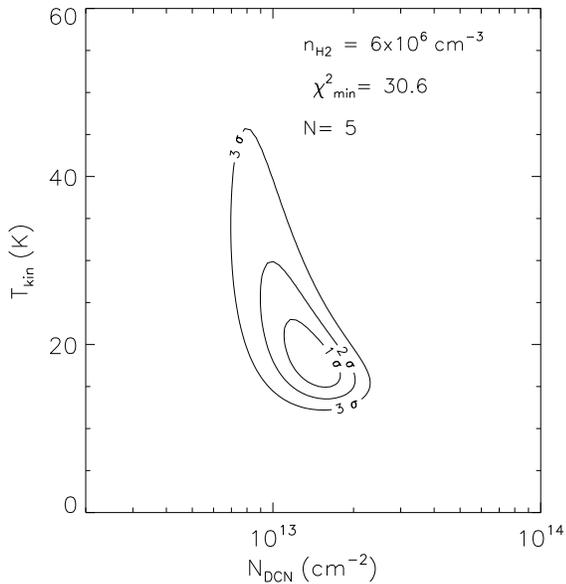}
\caption{LVG analysis results for the DCN molecule towards Clump 1. }
\label{dcn_lvg}
\end{figure}

Figure \ref{dcn_lvg} shows as an example the $\chi^2$ analysis for fitting LVG models for the DCN molecule towards clump 1. The $\chi^2$ was computed using the 4 DCN transitions, weighted by uncertainties including a 15\% calibration uncertainty of the integrated flux, as well as the opacity of the DCN(2-1) line as derived in \ref{section_hfs}.
The derived kinetic temperature is consistent within the 3$\sigma$ uncertainty range with the one derived from the methanol analysis. The DCN column density compares also very well with the rotational diagram result. The 1$\sigma$ confidence interval for the two parameters is shown in Table \ref{rottable}. We analyzed in the same way the DCN emission towards clump~3 as well as the H$^{13}$CN towards both clumps. The results are all shown in Table \ref{rottable}.

\subsection{Low DCO$^+$ emission}
\label{hco+}

\begin{figure}[!ht]
\includegraphics[width=10cm,angle=-90]{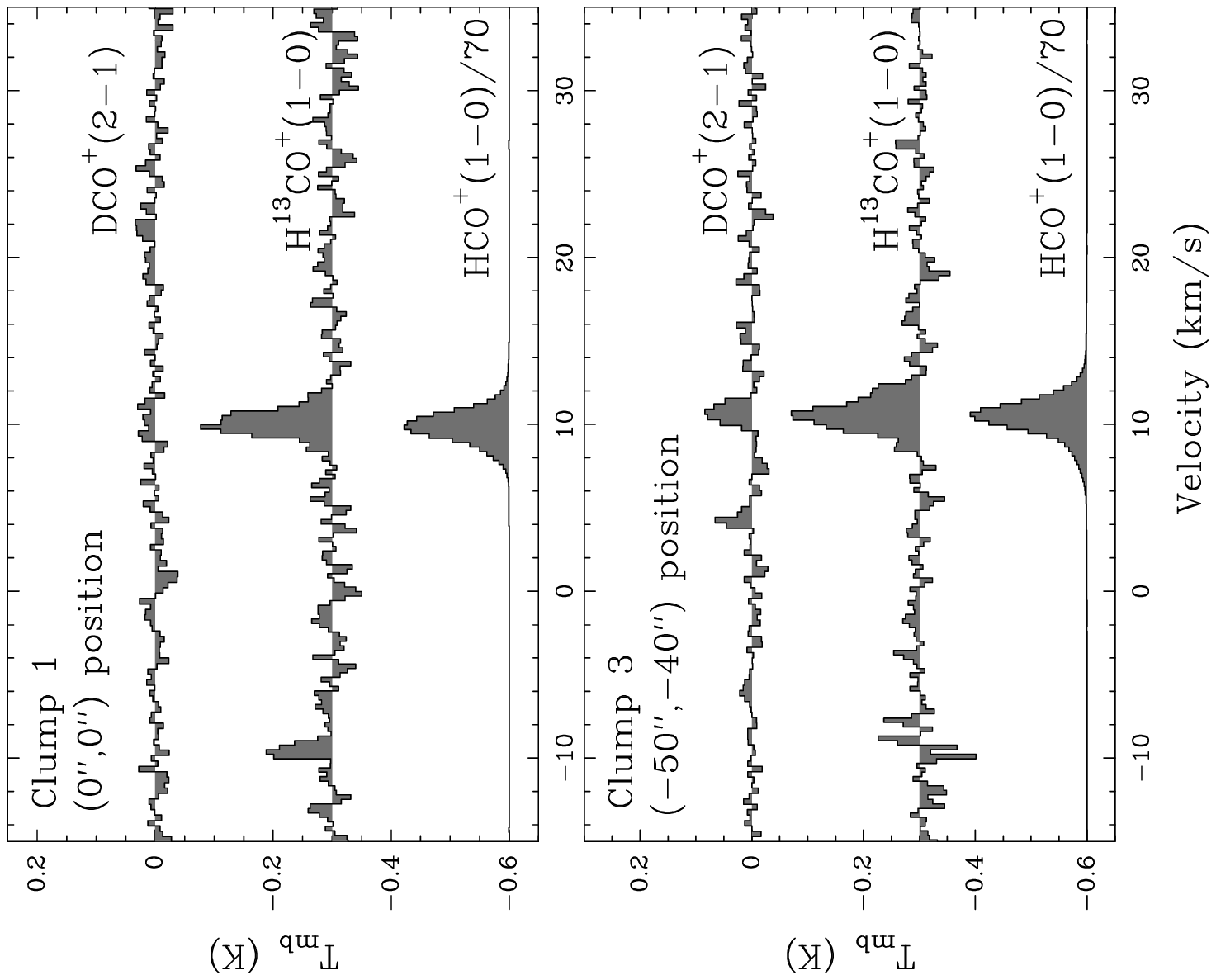}
\caption{DCO$^+$, H$^{13}$CO$^+$ and HCO$^+$ spectra. }
\label{dcop}
\end{figure}

DCO$^+$(2--1), H$^{13}$CO$^+$(1--0) and HCO$^+$(1--0) were observed towards the two clumps. The spectra are presented in Figure~\ref{dcop}. By comparison with H$^{13}$CO$^+$(1--0), HCO$^+$(1--0) is found to be optically thin. The observed parameters are listed in Table~\ref{flux}. Note that toward clump 3 the linewidth of the H$^{13}$CO$^+$ and HCO$^+$ lines are significantly larger than the HCN and DCN linewidths (as determined from the hyperfine structure fit). The thermal broadening expected for H$^{13}$CO$^+$ at 50\,K is 0.27\,km\,s$^{-1}$, and 0.47\,km\,s$^{-1}$ at 150\,K, the temperature characteristic of the interclump gas. This points to the fact that HCO$^+$ and H$^{13}$CO$^+$ are also present in the warmer interclump gas, in agreement with the interferometric studies of \citet{YoungOwl00} and \citet{Lis03}, who claimed that a large fraction of the emission of H$^{13}$CO$^+$(1-0) is extended and thus filtered out by the interferometer.

On the other hand, we do not expect deuterated molecules to be at all present in the hot interclump gas in steady-state. The fact that the main isotopomers are likely to trace both the two gas components implies that the D/H ratio derived from the observations should be considered as a lower limit. 

We estimate column densities towards the clumps (assuming resp. sizes of 7$''$ and 8$''$ for clump 1 and 3) for different rotational temperatures, assuming that all emission comes from the clumps. Although DCO$^+$ and H$^{13}$CO$^+$ column densities vary by a factor of two to three for temperatures between 20 and 70\,K, the abundance ratio is less affected by the temperature uncertainty (less than a factor of 1.5, see Table \ref{dcop_result})

\begin{table*}[!ht]
\caption{Rotational analysis for HCO$^+$ and isotopologues. Upper limits are 3$\sigma$. }
\begin{tabular}{lcccccc}
\noalign{\smallskip}
\hline
\hline
\noalign{\smallskip}
                        & \multicolumn{3}{c}{clump 1}   &  \multicolumn{3}{c}{clump 3} \\
T$_{\rm rot}$ &    N$_{\rm DCO^+}$  & N$_{\rm H^{13}CO^+}$ &  N$_{\rm DCO^+}$/N$_{\rm HCO^+}$$^1$ & N$_{\rm DCO^+}$  & N$_{\rm H^{13}CO^+}$ &  N$_{\rm DCO^+}$/N$_{\rm HCO^+}$$^1$  \\
           &    \small{(cm$^{-2}$)}  & \small{(cm$^{-2}$)}  & &  \small{(cm$^{-2}$)}  & \small{(cm$^{-2}$)}  & \\
\noalign{\smallskip}
\hline
\noalign{\smallskip}
20 K  &  $<$1.6$\times$10$^{11}$ & 1.0$\times$10$^{13}$ & $<$ 2.3$\times$10$^{-4}$ & (5.0$\pm$0.8)$\times$10$^{11}$ & (9.9$\pm$0.1)$\times$10$^{12}$ & (7.3$\pm$1.4)$\times$10$^{-4}$ \\
30 K  & $<$2.0$\times$10$^{11}$ &  1.4$\times$10$^{13}$  & $<$ 2.0$\times$10$^{-4}$ & (6.2$\pm$1.0)$\times$10$^{11}$ & (1.4$\pm$0.2)$\times$10$^{13}$ & (6.4$\pm$1.3)$\times$10$^{-4}$ \\
40 K  & $<$2.4$\times$10$^{11}$ & 1.8$\times$10$^{13}$  & $<$ 1.9$\times$10$^{-4}$ &  (7.5$\pm$1.2)$\times$10$^{11}$ & (1.8$\pm$0.2)$\times$10$^{13}$ & (6.0$\pm$1.1)$\times$10$^{-4}$ \\
50 K  &$<$2.8$\times$10$^{11}$ & 2.2$\times$10$^{13}$  & $<$ 1.8$\times$10$^{-4}$ &  (8.9$\pm$1.4)$\times$10$^{11}$ & (2.2$\pm$0.2)$\times$10$^{13}$ & (5.9$\pm$1.0)$\times$10$^{-4}$ \\
70 K  & $<$3.7$\times$10$^{11}$ &  3.1$\times$10$^{13}$  & $<$ 1.7$\times$10$^{-4}$ & (1.2$\pm$0.2)$\times$10$^{12}$ & (3.0$\pm$0.3)$\times$10$^{13}$ & (5.7$\pm$1.1)$\times$10$^{-4}$  \\
\noalign{\smallskip}
\hline
\noalign{\smallskip}
\end{tabular}\\
$^1$We assume HCO$^+$/H$^{13}$CO$^+$\,=\,70 \citep{Wilson99}.
\label{dcop_result}
\end{table*}

Can we quantify how much of the HCO$^+$ emission actually comes from the clumps? \citet{Hogerheijde95} modelled multi-frequency transitions of several molecules with a two-phase (interclump + clump) model, and they concluded that 10\% of the column density of the material is in the clumps. We can then get an estimate of the clump contribution to H$^{13}$CO$^+$ emission. Since the emission is likely dominated by the interclump gas, we estimate the column density of HCO$^+$ assuming a temperature of 150\,K. For clump 3, we then find a column density of H$^{13}$CO$^+$ of  6.2$\times$10$^{13}$ cm$^{-2}$. If the clump represents 10\% of the column density, then N(H$^{13}$CO$^+$) $\sim$ 6$\times$10$^{12}$ cm$^{-2}$ in the clump, which is roughly a factor of two to three lower than what we estimated. 
We may thus underestimate the DCO$^+$/ HCO$^+$ by a factor up to three.

\subsection{Detection of HDCO}

In order to test the possibility that the deuteration in the clumps originates in chemistry involving CH$_2$D$^+$, we targetted HDCO, a molecule synthesized in the gas phase via the CH$_2$D$^+$ route, through the following reactions \citep{Turner01,Roueff07}: \\
$$ {\rm CH_2D^+ + H_2 \rightarrow CH_4D^+   ~~~~~~~~~~(4) }$$ 
$$ {\rm CH_4D^+ + e^- \rightarrow CH_2D + H_2  ~~~~(5) } $$ 
$$ {\rm ~CH_2D + O \rightarrow HDCO + H  ~~~~~~~(6). } $$ 
 
\begin{figure}[!h]
\includegraphics[width=5.8cm, angle=-90]{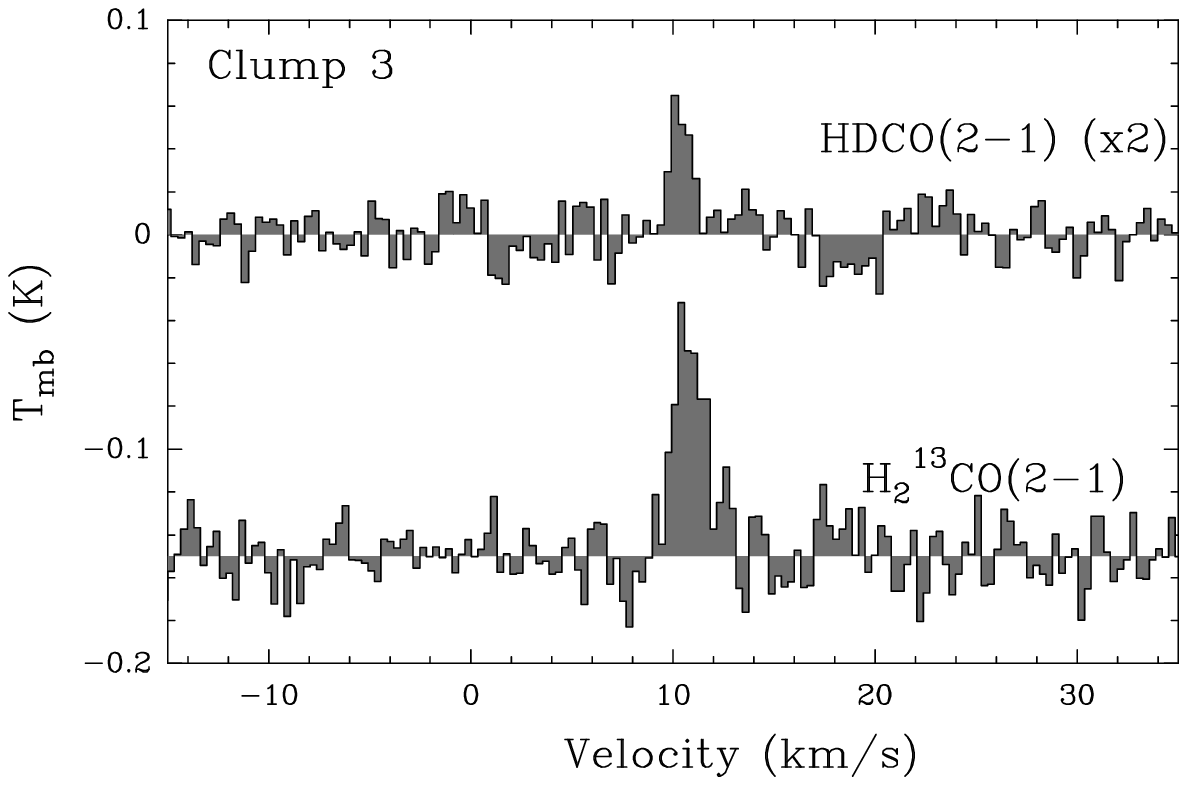}
\caption{HDCO and H$_2^{13}$CO detected towards Clump 3. }
\label{hdco}
\end{figure}

We detected the HDCO(2--1) transition towards clump\,3 (Fig. \ref{hdco}). The line has a V$_{\rm lsr}$ of 10.4\,km\,s$^{-1}$, consistent with other lines detected towards this clump. 
The line width (1.2\,km\,s$^{-1}$) is also similar to those of other deuterated species (DCN, DCO$^{+}$), making the detection of HDCO rather secure.
We also detected the H$_2$$^{13}$CO(2--1) line towards the same position. 

We computed the column density of the two molecules using the LTE approximation. The column densities vary by a factor of three depending on the assumed temperature between 20 and 70\,K. However, as the two observed lines have similar energies, the HDCO/H$_2$$^{13}$CO ratio is less sensitive to the temperature, varying only by a factor of 1.2 over the whole temperature range, which lies within the error bar of the determined ratio. We give in Table \ref{results} the column densities and D/H ratio for a fixed temperature of 35\,K (as pointed by the analysis of the CH$_3$OH emission). We assume again H$_2$$^{13}$CO/H$_2$CO = 70 \citep{Wilson99}.

One should note that H$_2$CO is also partly tracing the interclump gas (\citealt{Leurini06b}, Leurini et al. submitted). The derived HDCO/H$_2$CO ratio above is thus to be considered as a lower limit (see discussion in section \ref{compare_gp}).

\subsection{Upper limits on other interesting deuterated molecules} 

In the following, we derive upper limits on the fractional abundances of other deuterated molecules that were not detected. The discussion of the results is left to section \ref{discussion}.

\subsubsection{C$_2$D} 

C$_2$D is another molecule believed to form in the gas phase from CH$_2$D$^+$. \citet{Roueff07} predicted an D/H ratio for this molecule of 3.9$\times$10$^{-2}$, and an abundance of 6.6$\times$10$^{-11}$ with respect to H$_2$, in their low metal model at 50\,K.  

The J\,=\,2\,--\,1 band was observed towards clump 3. The VESPA correlator at the IRAM telescope showed unfortunately a lot of platforming. Using the 1MHz filter bank, the lines are not detected, and the rms is 8\,mK (T$_{mb}$ scale) at a resolution of 2.1\,km\,s$^{-1}$. Assuming a linewidth of 1.5 km\,s$^{-1}$ (as found in HCN and DCN observations), we derive the flux upper limit listed in Table \ref{flux}. Assuming a T$_{\rm rot}$ of 35\,K, this corresponds to an upper limit on the C$_2$D column density of 2.5\,$\times$\,10$^{13}$  cm$^{-2}$, and an upper limit for the fractional abundance of 2\,$\times$\,10$^{-10}$, compatible with the the prediction of \citet{Roueff07}.

\subsubsection{DNC}

DNC is synthesized in the gas phase, mainly from a route involving the H$_2$D$^+$ ion, as opposed to DCN which can be synthesized from CH$_2$D$^+$ \citep{Turner01, Roueff07}. We searched for the 2--1 transition of this species, to get constraints on the importance of the H$_2$D$^+$ chemistry in clump 3. We did not detect it, at a rms noise level of 9\,mK (T$_{mb}$ scale) and a resolution of 0.31\,km\,s$^{-1}$. We derive the flux upper limit listed in Table \ref{flux}. This corresponds to an upper limit of 1.5$\times$10$^{11}$\,cm$^{-3}$ for the DNC column density (assuming T$_{\rm rot}$ = 35\,K), and an upper limit on the abundance relative to H$_2$ of 1.0$\times$10$^{-12}$. Assuming that the HNC(4--3) line is optically thin, we find N$_{\rm HNC}$ = 1.05$\times$10$^{13}$ cm$^{-2}$. {\bf This has to be considered as a lower limit if HNC(4--3) is not optically thin.} This translates into an upper limit of DNC/HNC\,$<$\,1.4$\times$10$^{-2}$ (3$\sigma$). This upper limit is not very constraining compared to the deuteration ratios measured in HCN, because HNC is found to be more than three orders of magnitude less abundant than HCN. The very high HCN/HNC ratio in the clump is to our knowledge one of the highest observed so far. 

\subsubsection{CH$_2$DOH}

We searched for monodeuterated methanol, which is believed to trace the evaporation of highly deuterated ices \citep{Parise02,Parise04,Parise06a}. We looked for the 2$_{\rm K}$\,--\,1$_{\rm K}$ rotational band at 89\,GHz, corresponding to low energy transitions. CH$_2$DOH was not detected towards any of the two clumps. Rms levels of 4\,mK toward clump 1 and 6\,mK toward clump 3 (T$_{mb}$ scale) were reached, at a 0.26\,km/s resolution. Assuming a linewidth of 1.8 (resp. 1.5) \,km\,s$^{-1}$ for clump 1 (resp. 3), we derive the upper limit for the integrated intensity listed in Table \ref{flux}. This corresponds to an upper limit of 1.9\,$\times$\,10$^{14}$ cm$^{-2}$ on the CH$_2$DOH column density in clump 3, i.e. an upper limit on the fractional abundance of 1.5\,$\times$\,10$^{-9}$. 

\subsubsection{HDO}

The HDO line at 241\,GHz was searched for towards clump 3 but not detected, at a noise level of 47\,mK (rms, T$_{mb}$ scale), and a resolution of 0.3 km\,s$^{-1}$. 
The associated upper limit for the integrated line intensity given in Table \ref{flux} corresponds to an upper limit of 4.4\,$\times$\,10$^{13}$ cm$^{-2}$ for the HDO column density, i.e. an upper limit for the abundance of 3.4$\times$10$^{-10}$.

\begin{table*}[!ht]
\caption{Summary of column densities, abundances and D/H ratios in the observed molecules (assuming T$_{\rm rot}$=35\,K for species detected with only one transition).}
\begin{tabular}{lcccccc}
\noalign{\smallskip}
\hline
\hline
\noalign{\smallskip}
 & \multicolumn{3}{c}{clump 1}   &  \multicolumn{3}{c}{clump 3} \\
Molecule  & ~~~~~N & ~~~~~x & ~~~~~XD/XH &  ~~~~~N & ~~~~~x & ~~~~~XD/XH\\
           &    \small{(cm$^{-2}$)}  & \small{(cm$^{-3}$)}  & &  \small{(cm$^{-2}$)}  & \small{(cm$^{-3}$)}  & \\
\noalign{\smallskip}
\hline
\noalign{\smallskip}
H$^{13}$CN   & (3.1$\pm$0.4)$\times$10$^{13}$ &  1.9$\times$10$^{-10}$&  &(2.5$\pm$0.3)$\times$10$^{13}$ &  1.9$\times$10$^{-10}$&   \\
DCN     &    (1.4$\pm$0.3)$\times$10$^{13}$ &  8.8$\times$10$^{-11}$&  0.7$\pm$0.2 \%&  (1.9$\pm$0.3)$\times$10$^{13}$ &  1.5$\times$10$^{-10}$&  1.1$\pm$0.2 \%  \\
\noalign{\smallskip}
\hline
\noalign{\smallskip}
H$^{13}$CO$^+$ &  (2.0$\pm$1.0)$\times$10$^{13}$  &   1.3$\times$10$^{-10}$ & & (1.6$\pm$0.2)$\times$10$^{13}$ & 1.2$\times$10$^{-10}$ & \\
DCO$^+$    & $<$\,2.2$\times$10$^{11}$ &  $<$ 1.4$\times$10$^{-12}$ & $<$ 2$\times$10$^{-4}$  &  (6.9$\pm$1.1)$\times$10$^{11}$&   5.3$\times$10$^{-12}$ & (6.1$\pm$1.1) 10$^{-4}$\\
\noalign{\smallskip}
\hline
\noalign{\smallskip}
H$_2$$^{13}$CO   & -- & -- & -- & (1.2$\pm$0.1)$\times$10$^{13}$ &  9.2$\times$10$^{-11}$&  \\
HDCO   & -- & -- & -- &  (4.8$\pm$0.8)$\times$10$^{12}$&  3.7$\times$10$^{-11}$ &0.6$\pm$0.1 \% \\
\noalign{\smallskip}
\hline
\noalign{\smallskip}
C$_2$D  & -- & -- & -- &   $<$\,2.5$\times$10$^{13}$ &    $<$\,2$\times$10$^{-10}$   & -- \\
\noalign{\smallskip}
\hline
\noalign{\smallskip}
HNC  & -- & -- & -- & 1.1$\times$10$^{13}$ &  & \\
DNC  & -- & -- & -- & $<$\,1.5$\times$10$^{11}$ &  $<$\,1$\times$10$^{-12}$ & $<$\,1.4 \%\\
\noalign{\smallskip}
\hline
\noalign{\smallskip}
CH$_2$DOH & $<$\,1.7$\times$10$^{14}$ & $<$\,1.1$\times$10$^{-9}$ & -- & $<$\,1.9$\times$10$^{14}$ &   $<$\,1.5$\times$10$^{-9}$ & --\\
\noalign{\smallskip}
\hline
\noalign{\smallskip}
HDO & -- & -- & -- & $<$\,4.4$\times$10$^{13}$ &   $<$\,3.4$\times$10$^{-10}$ & --\\
\noalign{\smallskip}
\hline
\noalign{\smallskip}
\end{tabular}
\label{results}
\end{table*}

\section{Discussion}
\label{discussion}

The main results of this study are the confirmation through multi-transition observations of the significant deuterium fractionation of HCN found by \citet{Leurini06b}, as well as the detection of significant fractionation of HDCO in dense clumps in the Orion Bar PDR that are too warm for deuteration to be sustained by the H$_2$D$^+$ precursor. In the following, we consider the possible explanations for this high deuteration: products evaporated from ices surrounding dust grains, or gas-phase products. This study can also help understand the chemistry at work in PDRs.

\subsection{Grain evaporation ?}

One plausible origin for highly deuterated molecules is the evaporation of ices surrounding dust grains. Highly deuterated methanol, formaldehyde and water are observed in hot cores \citep[e.g. ][]{Jacq88, Comito03} and hot corinos \citep{Parise02, Parise04, Parise05a, Parise06a}, and are thought to be remnants of the cold prestellar phase.
 
In particular methanol is believed to form primarily on dust grains, as the gas-phase production routes for this molecule are not efficient. The same may also be true for water, the main constituent of ices surrounding dust grains. Deuterated isotopologues of these molecules may thus trace surface chemistry processes that enhance the deuterium content of these species, because the accreting atomic D/H ratio is much higher than the elemental D/H ($\sim$\,10$^{-5}$). Atomic accretion on dust grains and surface reactions take place in cold environment, i.e. in conditions which are also favoring a large atomic D/H. Although it is not clear which thermal history the gas in the Orion Bar has undergone in the past (are the clumps remnants of high density clumps of the cold molecular cloud, which are still shielded from the PDR radiation?), it may be possible that we are witnessing in the clumps evaporation of ices that were formed during the cold molecular cloud phase. In this case, we would expect the composition of the ices to be comparable to the ones evaporated in hot corinos, i.e. with high CH$_2$DOH/CH$_3$OH ratios.

The temperature of the grains should follow closely the gas temperature (typically $<$50\,K) in the clumps, due to the high density in the clumps \citep{Kruegel84}.  Although this temperature is high enough to sublimate CO ices (T$_{\rm evap}$$\sim$20\,K), it is not high enough to sublimate polar ices dominated by H$_2$O (T$_{\rm evap}$$\sim$100\,K). Pure HCN ices are expected to sublimate in the ISM conditions at temperatures $\sim$\,65\,K \citep[extrapolation from the sublimation temperatures in comets from][]{Prialnik06}. However, HCN is only a minor constituent of ices in objects where it was detected \citep[$<$\,3\% of water in W33A, 0.25\% of water in Hale Bopp,][]{Ehrenfreund00}. It will thus behave as small impurity in the water ice, and its sublimation will be dominated by the sublimation of H$_2$O. It is thus rather unlikely that evaporation plays an important role in the warm clumps, except maybe on the clump surfaces.

The low relative abundance of methanol (2$\times$10$^{-9}$) compared to the abundance in hot corinos \citep[$\sim$\,10$^{-7}$,][]{Maret05} may be a further sign that {\it thermal} evaporation is not the dominating chemical process, but rather that gas-phase chemistry is the most significant mechanism responsible for the enhanced deuteration levels. Note that grain chemistry models including non-thermal desorption of methanol are enough to explain these relatively low abundances of methanol \citep{Garrod07}. Of course it is also possible that evaporated ices get photodissociated in the PDR, and this effect will be studied in more detail in a forthcoming paper (Parise et al. in prep). We focus here now on an alternative explanation, assuming that gas phase reactions are more predominant.

\subsection{Gas-phase chemistry ?}

The low temperature of the clumps with respect to the sublimation temperature of water ices points to the fact that thermal ice evaporation does not play a predominant role in the clumps. At the same time, the grain temperature (which, at these high densities, should follow the observed gas temperature) is too high for CO to stick efficiently on dust grains, preventing efficient grain chemistry to take place. The present chemistry is thus likely dominated by gas-phase processes, and the clumps may be in this sense remnants of the molecular cloud, which have been warmed up to temperatures higher than for typical molecular clouds, due to their location behind the photoionising front. The CH$_3$OH abundance, in particular, is similar to that  measured towards the TMC-1 dark cloud \citep[3$\times$10$^{-9}$,][]{Smith04}, and this molecule appears to be a remnant of grain chemistry happening during the colder era of the cloud, followed by non-thermal desorbing processes \citep{Garrod07}. This possibility may be further tested by observing the D/H ratio in methanol. A formation of methanol during the cold molecular cloud phase is expected to lead to high CH$_2$DOH/CH$_3$OH ratios, as measured e.g. in prestellar cores \citep[between 5\% and 30\%][]{Bacmann07}. Unfortunately, the upper limit we derive here (CH$_2$DOH/CH$_3$OH\,$<$\,0.8) is not constraining enough, and further insight will require significantly deeper integrations, for which the sensitivity of ALMA will be needed.

For other molecules that can also form in the gas-phase, gas-phase processes are likely to be predominant. To test the hypothesis that gas-phase chemistry is the dominating process in the clumps, we compare in the remaining discussion the observed abundances and D/H ratios to the results of a pure steady-state gas-phase model. In particular, we are interested to see if high DCN/HCN ratio, low DCO$^+$/HCO$^+$ ratio and the detection of HDCO can be explained in the framework of a simple steady-state model.

\subsection{Comparison to gas-phase model}
\label{compare_gp}

We compare in the following the observed fractionation and relative abundances of molecules to the predictions of a pure gas-phase model, based on an updated version of the chemical model described by \citet{Roueff07}. The new model takes into account the new branching ratios of the N$_2$H$^+$ dissociative recombination \citep{Molek07}, computes the pre-exponential factors of the reverse reactions involved in the deuterium fractionation of CH$_3^+$  with the proper factors involved in the translational partition functions. In addition we have included the branching ratios of the electronic recombination of HCO$^+$  from \citet{Amano90} where the channel towards CO is found to be predominant. The radiative association reactions of CH$_3^+$ and deuterated substitues with H$_2$ have been derived from the theoretical predictions  of \citet{Bacchus00} who display values at different temperatures.

The exothermicity of the CH$_3^+$ + HD reaction and subsequent deuteration steps are not well constrained. These exothermicities were derived experimentally by \citet{Smith82a}. However, theoretical assessments from zero-point energies lead to higher barriers. This has the effect to allow deuteration to remain efficient at even higher temperatures (up to 70\,K). In the following, we will compare our observations to two models, calculated resp. by using the experimental values from \citet[][{\it saa} model]{Smith82a} and the exothermicities computed theoretically from zero-point energies ({\it theo} model). The respective exothermicities assumed in the models are listed in Tab. \ref{table_barrieres}.

\begin{table}
\caption{Exothermicities used in the two different chemical models. }
\begin{tabular}{lll}
\noalign{\smallskip}
\hline
\hline
\noalign{\smallskip}
Reaction & Exothermicity  & Exothermicity \\
& saa model & theo model \\
\noalign{\smallskip}
\hline
\noalign{\smallskip}
${\rm CH_3^+ + HD \Leftrightarrow CH_2D^+ + H_2}$ & 370 K& 670 K\\
${\rm CH_2D^+ + HD \Leftrightarrow CHD^+_2 + H_2}$ & 369 K& 433 K\\
${\rm CHD_2^+ + HD \Leftrightarrow CD_3^+ + H_2}$ & 379 K& 443 K\\
${\rm CH_3^+ + D_2 \Leftrightarrow CHD_2^+ + H_2}$ & 713 K& 1005 K\\
${\rm CH_3^+ + D_2 \Leftrightarrow CH_2D^+ + HD}$ & 319 K& 592 K\\
${\rm CH_2D^+ + D_2 \Leftrightarrow CD_3^+ + H_2}$ & 599 K& 564 K\\
${\rm CH_2D^+ + D_2 \Leftrightarrow CHD_2^+ + HD}$ & 317 K& 354 K\\
${\rm CHD_2^+  + D_2 \Leftrightarrow CD_3^+ + HD}$ & 290 K & 151 K\\
\noalign{\smallskip}
\hline
\noalign{\smallskip}
\end{tabular}
\label{table_barrieres}
\end{table}
 
\begin{figure*}[!ht]
\begin{tabular}{ll}
\includegraphics[width=8cm]{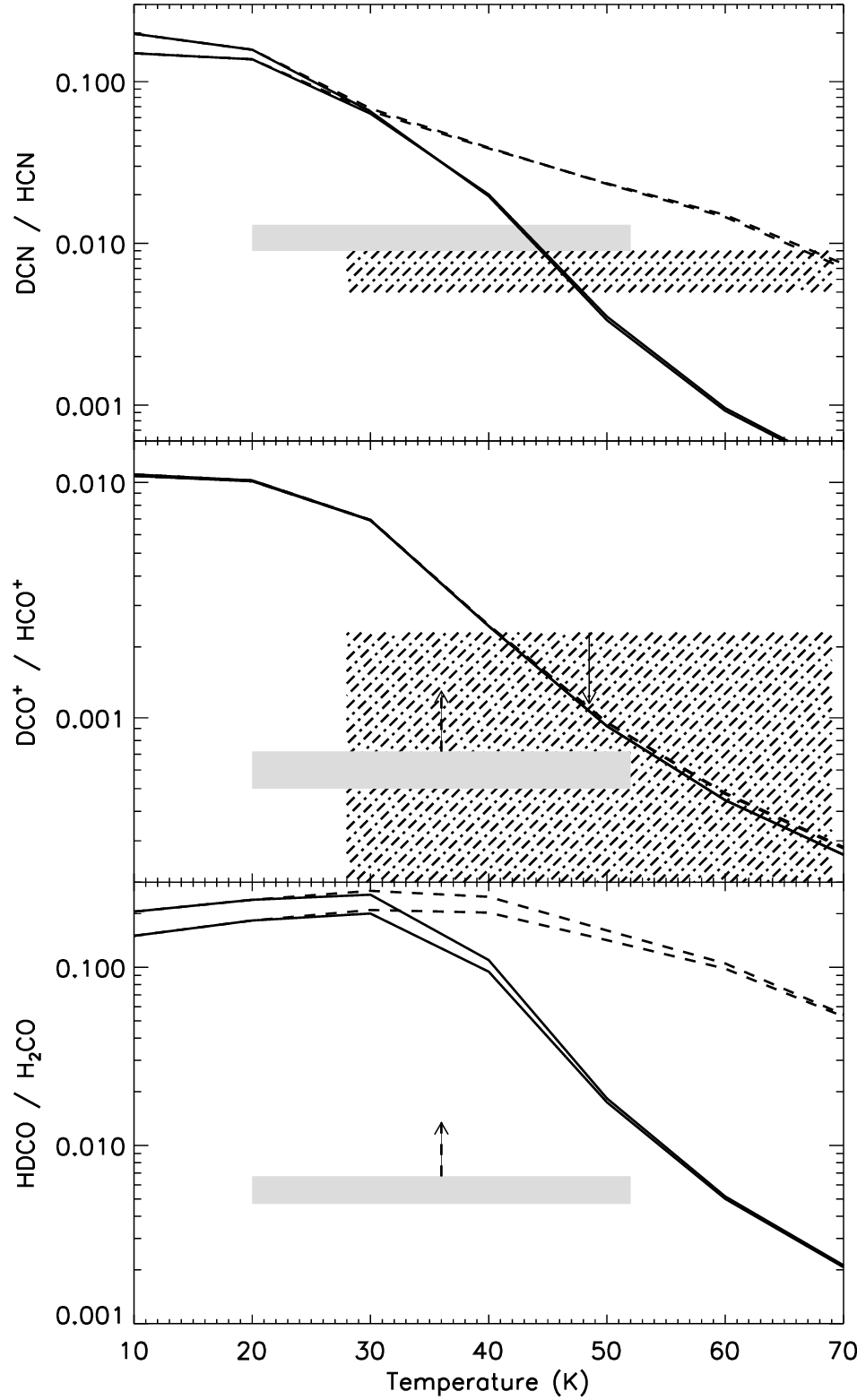} &
\includegraphics[width=8cm]{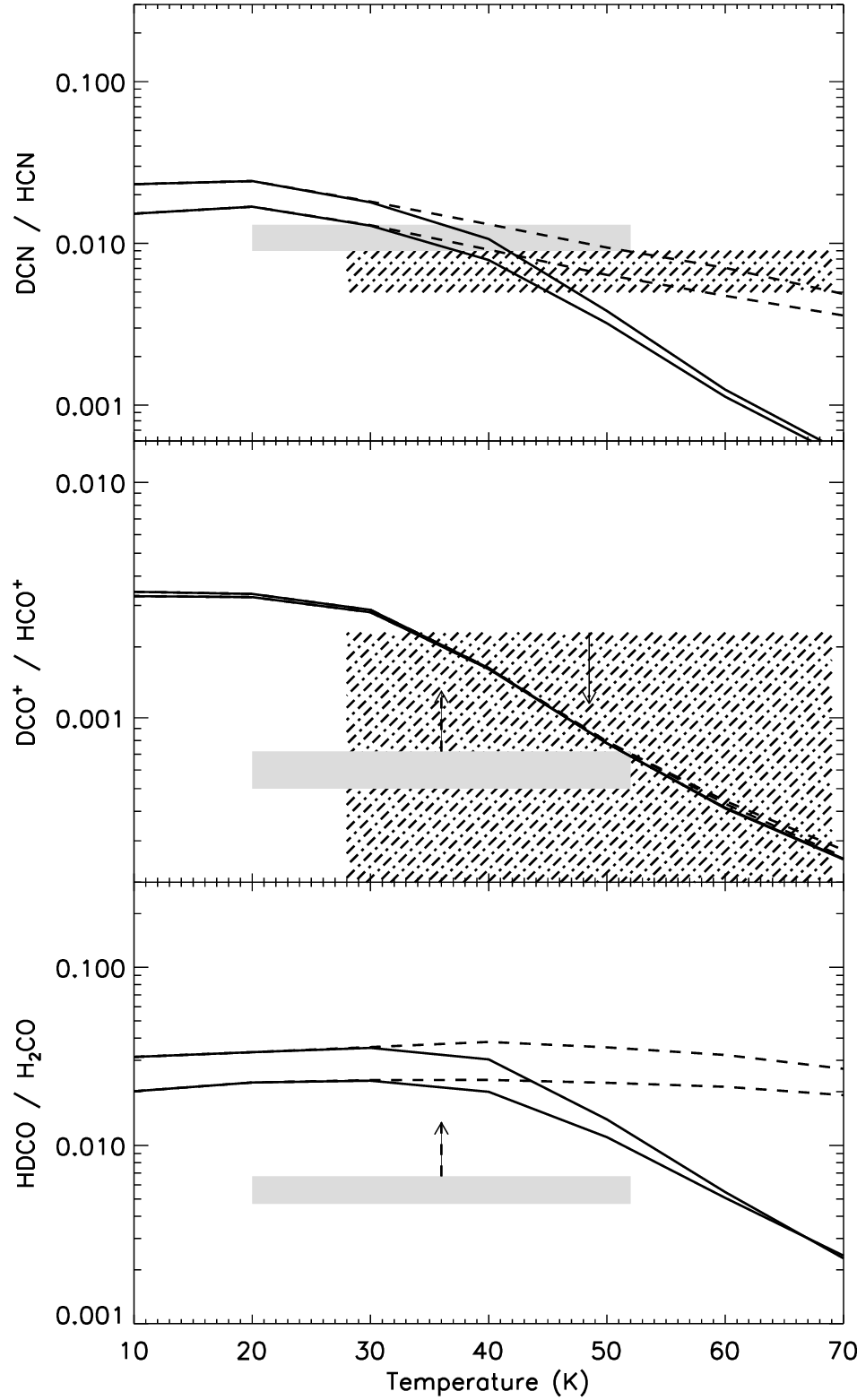} \\
\end{tabular}
\caption{Predictions of the D/H ratio for several molecules, as a function of the temperature, and observed ratios for clump 1 (dash-dot filling) and for clump 3 (grey filling). 
Model predictions from {\it saa} model (solid curves) and {\it theo} model (dashed curves) are computed 
for densities 3$\times$10$^6$ and 1$\times$10$^7$ cm$^{-3}$. Left panel: low-metal elemental abundances. Right panel: warm core elemental abundances \citep[see text and][]{Roueff07}. }
\label{compare_obs_model_clump3}
\end{figure*}

Figures \ref{compare_obs_model_clump1} and \ref{compare_obs_model_clump3} show the observed ratios towards clump 1 and 3 respectively, as well as the predicted ratios from the chemical models. The solid curves correspond to the {\it saa} model, and the dashed curves to the {\it theo} model. The two curves for each model correspond to the densities 3$\times$10$^6$ and 10$^7$ cm$^{-3}$ (in order of increasing D/H ratio). The grey filling delimitates the observed values (1$\sigma$) or upper limits (3$\sigma$), and the temperature range (3$\sigma$) as derived from the CH$_3$OH analysis. We computed the model predictions for the two different sets of elemental abundances used in \citet{Roueff07}. The ``warm core" elemental abundances are representative of a mostly undepleted gas, while the ``low metal" case, shown to lead to the best agreement to observations towards dense molecular clouds \citep{Graedel82}, involves moderate depletions of C, N and O, and strong depletions of S and Fe.

The models all show a decreasing fractionation with increasing temperature.
In the case of clump 1 (Figure \ref{compare_obs_model_clump1}), DCO$^+$ was not detected, and thus we have only an upper limit for  the observed DCO$^+$/HCO$^+$ ratio. The observed DCN/HCN ratio is consistent with the model prediction for a clump temperature of $\sim$\,45\,K, for the {\it saa} model with low-metallicity elemental abundances. The {\it theo}  model tends to overestimate the DCN/HCN ratio, requiring  temperatures $>$\,70\,K to reproduce the observed ratio, which are still consistent but rather at the high end of the uncertainty range of the temperature as determined from the methanol observations. Both models with warm core elemental abundances 
also qualitatively reproduce the DCN/HCN for a reasonable temperature range. The upper limit on the DCO$^+$/HCO$^+$ ratio is in the case of this clump not very constraining. Depending on the model, it points to T\,$>$\,33\,K or T\,$>$\,40\,K, which is consistent with the temperatures given by the DCN/HCN ratios.

\begin{figure*}[!ht]
\begin{tabular}{ll}
\includegraphics[width=8cm]{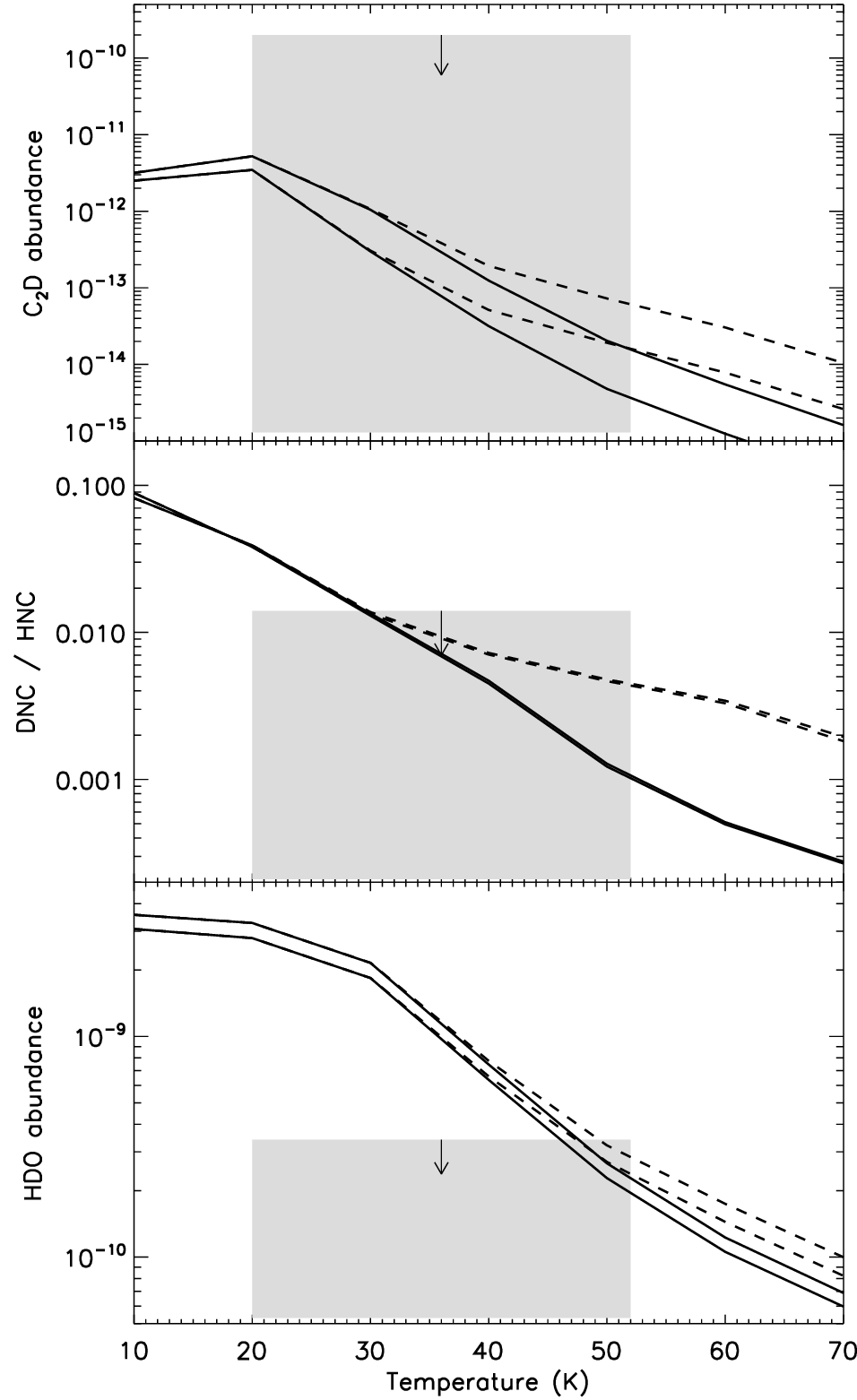} &
\includegraphics[width=8cm]{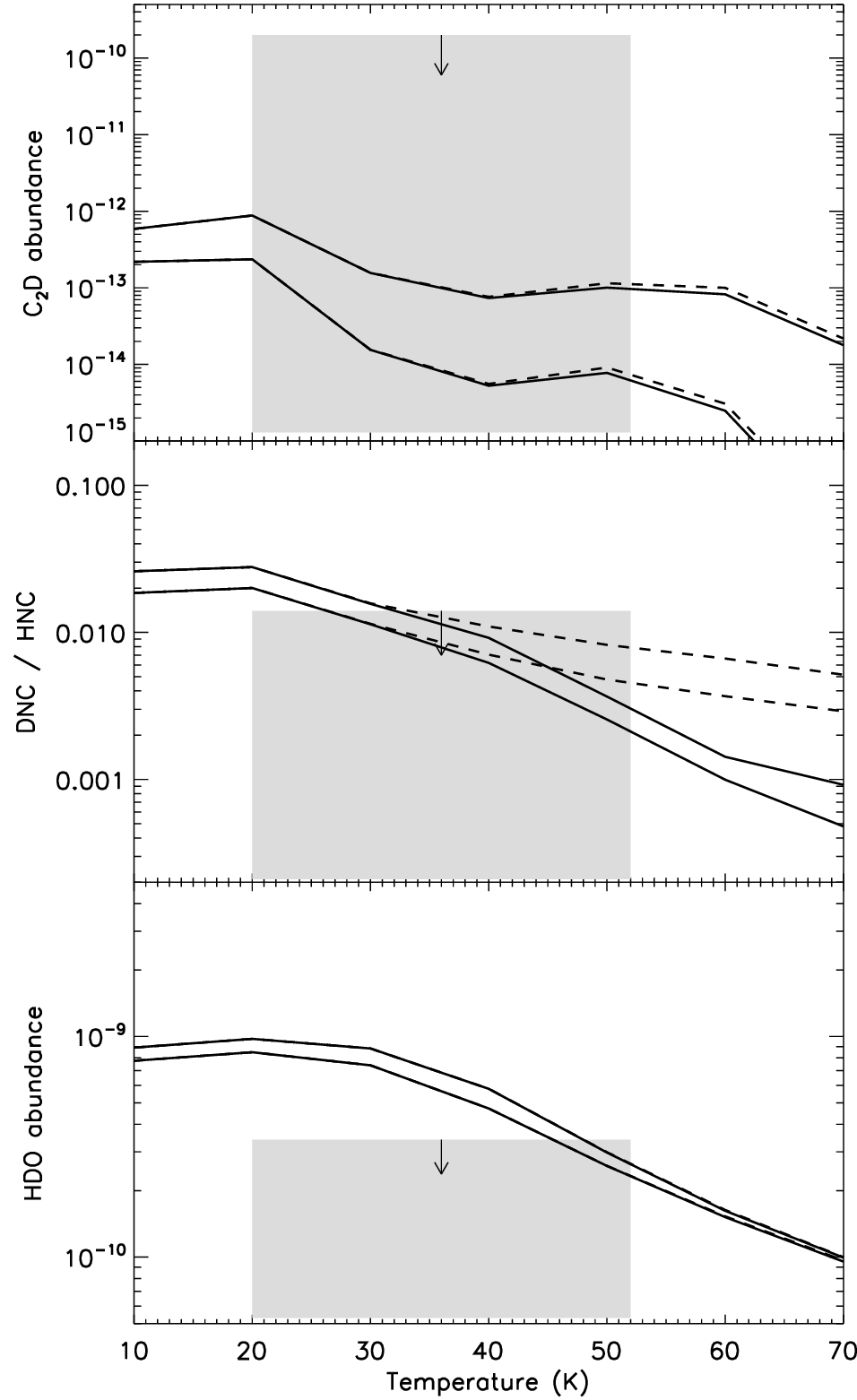} \\
\end{tabular}
\caption{Undetected molecules towards clump 3. Same plotting codes as for Fig. \ref{compare_obs_model_clump3}.}
\label{compare_obs_model_clump3_c2d}
\end{figure*}

In the case of clump 3 (Figure \ref{compare_obs_model_clump3}), both DCO$^+$/HCO$^+$ and HDCO/H$_2$CO ratios have been observed. The derived fractionation ratios have been estimated assuming that the main isotopologue is only tracing the clump. However, there is some evidence that H$_2$CO and HCO$^+$ are also present in the interclump medium. The case for HCO$^+$ was discussed in section \ref{hco+}, where we argued that we may have underestimated the DCO$^+$/HCO$^+$ ratio by up to a factor of three. 
H$_2$CO has been shown to be roughly as abundant in the clumps as in the interclump gas (Leurini et al. submitted). They derive an H$_2$CO column density of 6$\times$10$^{13}$ to 6$\times$10$^{14}$ cm$^{-3}$ in the interclump gas, and of (2.8--5.4) 10$^{14}$ cm$^{-3}$ towards clump 1, which is up to a factor of three lower than our estimate from H$_2^{13}$CO towards clump 3. We may thus be overestimating the H$_2$CO abundance in clump 3 by a factor of two to three. 
 As a consequence, the ratios we have derived are in fact only to be considered as lower limits for  the true ratios in the clumps (represented by the ascending arrows on Fig. \ref{compare_obs_model_clump3}). Note that this is not the case for the DCN/HCN ratio, because H$^{13}$CN has been shown to trace only the clumps \citep{Lis03}. In the case of clump 3, the DCN/HCN ratio points to a temperature of $\sim$45\,K (low-metal elemental abundances),  or 20-50\,K (warm core conditions). Both DCO$^+$/HCO$^+$ and HDCO/H$_2$CO lower limits are in principle consistent with the model. However, in the case of the low-metal {\it saa} model, for a temperature of 45\,K, agreement with the model would require that we increase the H$_2$CO abundance in the clump by a factor of six, which seems not consistent with the study of Leurini et al. (submitted). For this clump, better qualitative agreement is found with the models with warm core elemental conditions. 
It is also instructive to see if the upper limits on other deuterated molecules (C$_2$D, DNC, HDO) are consistent with the models (Figure  \ref{compare_obs_model_clump3_c2d}).  In the case of HDO and C$_2$D, where we have no observations of the main isotopologue, we compare the abundances directly with the model predictions. C$_2$D upper limits are not constraining ---a two orders of magnitude increase in  the sensitivity would be required to add interesting constraints. The DNC/HNC upper limits point to T\,$>$\,30\,K. But the most constraining molecule is certainly HDO,  which non-detection is consistent with the models only for T$>$ 40-50\,K. The fact that the HDO abundance seems to be consistent and even rather on the lower side of the model predictions is certainly a good hint that {\it thermal} evaporation is not playing a dominant role in injecting water molecules into the gas-phase.

As a conclusion from this comparison, we find a good qualitative agreement between the observations and pure steady-state gas-phase models. Although such a comparison is obviously limited, because the physical structure (temperature and density gradient) of the clump is not taken into account, and because the chemistry is only computed at steady-state, it shows that the scenario of warm gas phase deuterium chemistry is  viable for explaining the high deuteration ratios observed in DCN towards the Orion Bar clumps.

The comparison with the models is unfortunately limited by our poor handle on the clump temperature. Good probes of the temperature are usually the inversion lines of the NH$_3$ molecule. NH$_3$ was mapped in the Orion Bar by \citet{Batrla03} with single-dish. Although they find that the NH$_3$ emission comes mostly from the same regions as the HCN emission, they find high kinetic temperatures (T\,$>$\,100\,K) and argue that ammonia is located in the surface layers of the clumps, where icy mantle around dust grains evaporate. The traced temperature is thus rather typical of that of the interclump gas, and NH$_3$ is therefore not a good probe of the clump temperature. Deeper observations of several lines of CH$_3$OH, including high-J transitions might allow to constrain the temperature better.  

On the model side, a more detail treatment taking into account photodissociation in the PDR is the next step to take.

\section{Conclusions}

We have presented observations of deuterated molecules towards two dense clumps in the Orion Bar photodissociation region. These observations were aimed at confirming and understanding the origin of the DCN emission first detected in this region by \citet{Leurini06b}. We confirmed through the observation of four transitions of DCN the detection of this molecule towards one clump, and detected it towards a second clump of the Orion Bar. We also detected DCO$^+$ and HDCO towards this second clump, and provided upper limits for the abundance of other relevant deuterated molecules.
From the observation of these several species, formed by chemistry induced either by H$_2$D$^+$ or CH$_2$D$^+$, we find evidence based on a pure gas-phase chemistry model that the main ion responsible for deuteron transfer in the Orion Bar is CH$_2$D$^+$, as opposed to previously observed cases of colder regions or hot cores where H$_2$D$^+$ was the main actor (in the case of hot cores, deuterium fractionation is believed to be a fossil of cold chemistry in the earlier cold evolutionary phase, preserved into ice mantles). The luke-warm conditions in the Orion Bar clumps thus allowed us to observationally test chemical models in a different temperature range than most previous studies dealing with deuterium fractionation.  More refined understanding of the chemistry at work in the Orion Bar will require more detailed chemical modelling, coupling PDR models with gas-grain chemical networks.

\begin{acknowledgements}
We are very grateful to the APEX staff, in particular P. Bergman and A. Lundgren, for performing part of the APEX observations presented here. B.P. thanks M. Wernli for providing the HCN-He collision rates before publication, and for enlightening discussions. B.P. acknowledges fruitful discussions with A.G.G.M. Tielens and R. Garrod, and enlightening comments on statistics from Edward Polehampton and Didier Pelat. This work was funded by a Alexander von Humboldt research fellowship and by a Deutsche Forschungsgemeinschaft Emmy Noether project. D.C. Lis is supported by the U.S. National Science Foundation, award AST-0540882 to the Caltech Submillimeter Observatory.
\end{acknowledgements}

\bibliographystyle{aa}
\bibliography{/Users/bparise/These/Manuscrit/biblio}

\begin{thebibliography}{63}
\expandafter\ifx\csname natexlab\endcsname\relax\def\natexlab#1{#1}\fi

\bibitem[{{Amano}(1990)}]{Amano90}
{Amano}, T. 1990, J. Ch. Ph., 92, 6492

\bibitem[{{Asvany} {et~al.}(2004){Asvany}, {Schlemmer}, \&
  {Gerlich}}]{Asvany04}
{Asvany}, O., {Schlemmer}, S., \& {Gerlich}, D. 2004, \apj, 617, 685

\bibitem[{{Bacchus-Montabonel} {et~al.}(2000){Bacchus-Montabonel}, {Talbi}, \&
  {Persico}}]{Bacchus00}
{Bacchus-Montabonel}, M., {Talbi}, D., \& {Persico}, M. 2000, J. Phys. B, 33,
  955

\bibitem[{{Bacmann} {et~al.}(2007){Bacmann}, {Lefloch}, {Parise}, {Ceccarelli},
  \& {Steinacker}}]{Bacmann07}
{Bacmann}, A., {Lefloch}, B., {Parise}, B., {Ceccarelli}, C., \& {Steinacker},
  J. 2007, in Molecules in Space and Laboratory

\bibitem[{{Batrla} \& {Wilson}(2003)}]{Batrla03}
{Batrla}, W. \& {Wilson}, T.~L. 2003, \aap, 408, 231

\bibitem[{{Caselli} {et~al.}(2003){Caselli}, {van der Tak}, {Ceccarelli}, \&
  {Bacmann}}]{Caselli03}
{Caselli}, P., {van der Tak}, F.~F.~S., {Ceccarelli}, C., \& {Bacmann}, A.
  2003, \aap, 403, L37

\bibitem[{{Caselli} {et~al.}(2008){Caselli}, {Vastel}, {Ceccarelli}, {van der
  Tak}, {Crapsi}, \& {Bacmann}}]{Caselli08}
{Caselli}, P., {Vastel}, C., {Ceccarelli}, C., {et~al.} 2008, \aap, 492, 703

\bibitem[{{Comito} {et~al.}(2003){Comito}, {Schilke}, {G\'erin}, {Phillips},
  {Zmuidzinas}, \& {Lis}}]{Comito03}
{Comito}, C., {Schilke}, P., {G\'erin}, M., {et~al.} 2003, \aap, 402, 635

\bibitem[{{Ehrenfreund} \& {Charnley}(2000)}]{Ehrenfreund00}
{Ehrenfreund}, P. \& {Charnley}, S.~B. 2000, \araa, 38, 427

\bibitem[{{Flower} {et~al.}(2004){Flower}, {Pineau des For{\^e}ts}, \&
  {Walmsley}}]{Flower04}
{Flower}, D.~R., {Pineau des For{\^e}ts}, G., \& {Walmsley}, C.~M. 2004, \aap,
  427, 887

\bibitem[{{Garrod} {et~al.}(2007){Garrod}, {Wakelam}, \& {Herbst}}]{Garrod07}
{Garrod}, R.~T., {Wakelam}, V., \& {Herbst}, E. 2007, \aap, 467, 1103

\bibitem[{{Gerlich} {et~al.}(2002){Gerlich}, {Herbst}, \& {Roueff}}]{Gerlich02}
{Gerlich}, D., {Herbst}, E., \& {Roueff}, E. 2002, \planss, 50, 1275

\bibitem[{{Graedel} {et~al.}(1982){Graedel}, {Langer}, \&
  {Frerking}}]{Graedel82}
{Graedel}, T.~E., {Langer}, W.~D., \& {Frerking}, M.~A. 1982, \apjs, 48, 321

\bibitem[{{G{\"u}sten} {et~al.}(2006){G{\"u}sten}, {Nyman}, {Schilke},
  {Menten}, {Cesarsky}, \& {Booth}}]{Guesten06}
{G{\"u}sten}, R., {Nyman}, L.~{\AA}., {Schilke}, P., {et~al.} 2006, \aap, 454,
  L13

\bibitem[{{Hatchell} {et~al.}(1998){Hatchell}, {Millar}, \&
  {Rodgers}}]{Hatchell98}
{Hatchell}, J., {Millar}, T.~J., \& {Rodgers}, S.~D. 1998, \aap, 332, 695

\bibitem[{{Herbst} {et~al.}(1987){Herbst}, {Adams}, {Smith}, \&
  {Defrees}}]{Herbst87}
{Herbst}, E., {Adams}, N.~G., {Smith}, D., \& {Defrees}, D.~J. 1987, \apj, 312,
  351

\bibitem[{{Hogerheijde} {et~al.}(1995){Hogerheijde}, {Jansen}, \& {van
  Dishoeck}}]{Hogerheijde95}
{Hogerheijde}, M.~R., {Jansen}, D.~J., \& {van Dishoeck}, E.~F. 1995, \aap,
  294, 792

\bibitem[{{Huntress}(1977)}]{Huntress77}
{Huntress}, W.~T. 1977, \apjs, 33, 495

\bibitem[{{Jacq} {et~al.}(1988){Jacq}, {Henkel}, {Walmsley}, {Jewell}, \&
  {Baudry}}]{Jacq88}
{Jacq}, T., {Henkel}, C., {Walmsley}, C.~M., {Jewell}, P.~R., \& {Baudry}, A.
  1988, \aap, 199, L5

\bibitem[{{Jansen} {et~al.}(1995){Jansen}, {Spaans}, {Hogerheijde}, \& {van
  Dishoeck}}]{Jansen95}
{Jansen}, D.~J., {Spaans}, M., {Hogerheijde}, M.~R., \& {van Dishoeck}, E.~F.
  1995, \aap, 303, 541

\bibitem[{{Klein} {et~al.}(2006){Klein}, {Philipp}, {Kr{\"a}mer}, {Kasemann},
  {G{\"u}sten}, \& {Menten}}]{Klein06}
{Klein}, B., {Philipp}, S.~D., {Kr{\"a}mer}, I., {et~al.} 2006, \aap, 454, L29

\bibitem[{{Kruegel} \& {Walmsley}(1984)}]{Kruegel84}
{Kruegel}, E. \& {Walmsley}, C.~M. 1984, \aap, 130, 5

\bibitem[{{Leurini} {et~al.}(2009){Leurini}, {Parise}, {Schilke}, {Pety}, \&
  {Rolffs}}]{Leurini09}
{Leurini}, S., {Parise}, B., {Schilke}, P., {Pety}, J., \& {Rolffs}, R. 2009,
  \aap, subm.

\bibitem[{{Leurini} {et~al.}(2006){Leurini}, {Rolffs}, {Thorwirth}, {Parise},
  {Schilke}, {Comito}, {Wyrowski}, {G{\"u}sten}, {Bergman}, {Menten}, \&
  {Nyman}}]{Leurini06b}
{Leurini}, S., {Rolffs}, R., {Thorwirth}, S., {et~al.} 2006, \aap, 454, L47

\bibitem[{{Leurini} {et~al.}(2004){Leurini}, {Schilke}, {Menten}, {Flower},
  {Pottage}, \& {Xu}}]{Leurini04}
{Leurini}, S., {Schilke}, P., {Menten}, K.~M., {et~al.} 2004, \aap, 422, 573

\bibitem[{{Leurini} {et~al.}(2007){Leurini}, {Schilke}, {Wyrowski}, \&
  {Menten}}]{Leurini07}
{Leurini}, S., {Schilke}, P., {Wyrowski}, F., \& {Menten}, K.~M. 2007, \aap,
  466, 215

\bibitem[{{Linsky}(2003)}]{Linsky03}
{Linsky}, J.~L. 2003, Space Science Reviews, 106, 49

\bibitem[{{Lique} {et~al.}(2008){Lique}, {Tobo{\l}a}, {K{\l}os}, {Feautrier},
  {Spielfiedel}, {Vincent}, {Cha{\l}asi{\'n}ski}, \& {Alexander}}]{Lique08}
{Lique}, F., {Tobo{\l}a}, R., {K{\l}os}, J., {et~al.} 2008, \aap, 478, 567

\bibitem[{{Lis} {et~al.}(2002){Lis}, {Roueff}, {Gérin}, {Phillips}, {Coudert},
  {van der Tak}, \& {Schilke}}]{Lis02}
{Lis}, D.~C., {Roueff}, E., {Gérin}, M., {et~al.} 2002, \apjl, 571, L55

\bibitem[{{Lis} \& {Schilke}(2003)}]{Lis03}
{Lis}, D.~C. \& {Schilke}, P. 2003, \apjl, 597, L145

\bibitem[{{M{\" u}ller} {et~al.}(2001){M{\" u}ller}, {Thorwirth}, {Roth}, \&
  {Winnewisser}}]{Muller01}
{M{\" u}ller}, H.~S.~P., {Thorwirth}, S., {Roth}, D.~A., \& {Winnewisser}, G.
  2001, \aap, 370, L49

\bibitem[{{Maret} {et~al.}(2005){Maret}, {Ceccarelli}, {Tielens}, {Caux},
  {Lefloch}, {Faure}, {Castets}, \& {Flower}}]{Maret05}
{Maret}, S., {Ceccarelli}, C., {Tielens}, A.~G.~G.~M., {et~al.} 2005, \aap,
  442, 527

\bibitem[{{Maret} {et~al.}(2009){Maret}, {Faure}, {Scifoni}, \&
  {Wiesenfeld}}]{Maret09}
{Maret}, S., {Faure}, A., {Scifoni}, E., \& {Wiesenfeld}, L. 2009, ArXiv
  e-prints

\bibitem[{{Menten} {et~al.}(2007){Menten}, {Reid}, {Forbrich}, \&
  {Brunthaler}}]{Menten07}
{Menten}, K.~M., {Reid}, M.~J., {Forbrich}, J., \& {Brunthaler}, A. 2007, \aap,
  474, 515

\bibitem[{Molek {et~al.}(2007)Molek, McLain, Poterya, \& Adams}]{Molek07}
Molek, C., McLain, J., Poterya, V., \& Adams, N.~G. 2007, J. Phys. Chem. A,
  111, 6760

\bibitem[{{M{\"u}ller} {et~al.}(2005){M{\"u}ller}, {Schl{\"o}der}, {Stutzki},
  {Schlemmer}, {Giesen}, \& {Schilke}}]{Muller05}
{M{\"u}ller}, H.~S.~P., {Schl{\"o}der}, F., {Stutzki}, J., {et~al.} 2005, in
  IAU Symposium, ed. D.~C. {Lis}, G.~A. {Blake}, \& E.~{Herbst}, 30--+

\bibitem[{{Ossenkopf} \& {Henning}(1994)}]{Ossenkopf94}
{Ossenkopf}, V. \& {Henning}, T. 1994, \aap, 291, 943

\bibitem[{{Pagani} {et~al.}(2009){Pagani}, {Vastel}, {Hugo}, {Kokoouline},
  {Greene}, {Bacmann}, {Bayet}, {Ceccarelli}, {Peng}, \&
  {Schlemmer}}]{Pagani09}
{Pagani}, L., {Vastel}, C., {Hugo}, E., {et~al.} 2009, \aap, 494, 623

\bibitem[{{Parise} {et~al.}(2004){Parise}, {Castets}, {Herbst}, {Caux},
  {Ceccarelli}, {Mukhopadhyay}, \& {Tielens}}]{Parise04}
{Parise}, B., {Castets}, A., {Herbst}, E., {et~al.} 2004, \aap, 416, 159

\bibitem[{{Parise} {et~al.}(2005){Parise}, {Caux}, {Castets}, {Ceccarelli},
  {Loinard}, {Tielens}, {Bacmann}, {Cazaux}, {Comito}, {Helmich}, {Kahane},
  {Schilke}, {van Dishoeck}, {Wakelam}, \& {Walters}}]{Parise05a}
{Parise}, B., {Caux}, E., {Castets}, A., {et~al.} 2005, \aap, 431, 547

\bibitem[{{Parise} {et~al.}(2006){Parise}, {Ceccarelli}, {Tielens}, {Castets},
  {Caux}, {Lefloch}, \& {Maret}}]{Parise06a}
{Parise}, B., {Ceccarelli}, C., {Tielens}, A.~G.~G.~M., {et~al.} 2006, \aap,
  453, 949

\bibitem[{{Parise} {et~al.}(2002){Parise}, {Ceccarelli}, {Tielens}, {Herbst},
  {Lefloch}, {Caux}, {Castets}, {Mukhopadhyay}, {Pagani}, \&
  {Loinard}}]{Parise02}
{Parise}, B., {Ceccarelli}, C., {Tielens}, A.~G.~G.~M., {et~al.} 2002, \aap,
  393, L49

\bibitem[{{Pety} {et~al.}(2007){Pety}, {Goicoechea}, {Hily-Blant}, {Gerin}, \&
  {Teyssier}}]{Pety07}
{Pety}, J., {Goicoechea}, J.~R., {Hily-Blant}, P., {Gerin}, M., \& {Teyssier},
  D. 2007, \aap, 464, L41

\bibitem[{{Phillips} {et~al.}(1996){Phillips}, {Maluendes}, \&
  {Green}}]{Phillips96}
{Phillips}, T.~R., {Maluendes}, S., \& {Green}, S. 1996, \apjs, 107, 467

\bibitem[{{Pottage} {et~al.}(2002){Pottage}, {Flower}, \& {Davis}}]{Pottage02}
{Pottage}, J.~T., {Flower}, D.~R., \& {Davis}, S.~L. 2002, Journal of Physics B
  Atomic Molecular Physics, 35, 2541

\bibitem[{{Pottage} {et~al.}(2004){Pottage}, {Flower}, \& {Davis}}]{Pottage04}
{Pottage}, J.~T., {Flower}, D.~R., \& {Davis}, S.~L. 2004, Journal of Physics B
  Atomic Molecular Physics, 37, 165

\bibitem[{{Prialnik}(2006)}]{Prialnik06}
{Prialnik}, D. 2006, in IAU Symposium, Vol. 229, Asteroids, Comets, Meteors,
  ed. L.~{Daniela}, M.~{Sylvio Ferraz}, \& F.~J. {Angel}, 153--170

\bibitem[{{Risacher} {et~al.}(2006){Risacher}, {Vassilev}, {Monje}, {Lapkin},
  {Belitsky}, {Pavolotsky}, {Pantaleev}, {Bergman}, {Ferm}, {Sundin},
  {Svensson}, {Fredrixon}, {Meledin}, {Gunnarsson}, {Hagstr{\"o}m},
  {Johansson}, {Olberg}, {Booth}, {Olofsson}, \& {Nyman}}]{Risacher06}
{Risacher}, C., {Vassilev}, V., {Monje}, R., {et~al.} 2006, \aap, 454, L17

\bibitem[{{Roberts} {et~al.}(2003){Roberts}, {Herbst}, \& {Millar}}]{Roberts03}
{Roberts}, H., {Herbst}, E., \& {Millar}, T.~J. 2003, \apjl, 591, L41

\bibitem[{{Roueff} {et~al.}(2007){Roueff}, {Parise}, \& {Herbst}}]{Roueff07}
{Roueff}, E., {Parise}, B., \& {Herbst}, E. 2007, \aap, 464, 245

\bibitem[{{Schilke} {et~al.}(2001){Schilke}, {Pineau des For{\^e}ts},
  {Walmsley}, \& {Mart{\'{\i}}n-Pintado}}]{Schilke01}
{Schilke}, P., {Pineau des For{\^e}ts}, G., {Walmsley}, C.~M., \&
  {Mart{\'{\i}}n-Pintado}, J. 2001, \aap, 372, 291

\bibitem[{{Schilke} {et~al.}(1992){Schilke}, {Walmsley}, {Pineau des For\^ets},
  {Roueff}, {Flower}, \& {Guilloteau}}]{Schilke92}
{Schilke}, P., {Walmsley}, C.~M., {Pineau des For\^ets}, G., {et~al.} 1992,
  \aap, 256, 595

\bibitem[{{Smith} {et~al.}(1982){Smith}, {Adams}, \& {Alge}}]{Smith82a}
{Smith}, D., {Adams}, N.~G., \& {Alge}, E. 1982, \jcp, 77, 1261

\bibitem[{{Smith} {et~al.}(2004){Smith}, {Herbst}, \& {Chang}}]{Smith04}
{Smith}, I.~W.~M., {Herbst}, E., \& {Chang}, Q. 2004, \mnras, 350, 323

\bibitem[{{Turner}(2001)}]{Turner01}
{Turner}, B.~E. 2001, \apjs, 136, 579

\bibitem[{{van der Tak} {et~al.}(2002){van der Tak}, {Schilke}, {M{\" u}ller},
  {Lis}, {Phillips}, {Gérin}, \& {Roueff}}]{vanderTak02}
{van der Tak}, F.~F.~S., {Schilke}, P., {M{\" u}ller}, H.~S.~P., {et~al.} 2002,
  \aap, 388, L53

\bibitem[{{Walmsley} {et~al.}(2004){Walmsley}, {Flower}, \& {Pineau des For{\^
  e}ts}}]{Walmsley04}
{Walmsley}, C.~M., {Flower}, D.~R., \& {Pineau des For{\^ e}ts}, G. 2004, \aap,
  418, 1035

\bibitem[{{Wernli} \& {Faure}(2009)}]{Wernli09}
{Wernli}, M. \& {Faure}, A. 2009, MNRAS, in prep

\bibitem[{{Wernli} {et~al.}(2006){Wernli}, {Valiron}, {Faure}, {Wiesenfeld},
  {Jankowski}, \& {Szalewicz}}]{Wernli06}
{Wernli}, M., {Valiron}, P., {Faure}, A., {et~al.} 2006, \aap, 446, 367

\bibitem[{{Wernli} {et~al.}(2007){Wernli}, {Wiesenfeld}, {Faure}, \&
  {Valiron}}]{Wernli07}
{Wernli}, M., {Wiesenfeld}, L., {Faure}, A., \& {Valiron}, P. 2007, \aap, 464,
  1147

\bibitem[{{Wilson}(1999)}]{Wilson99}
{Wilson}, T.~L. 1999, Reports of Progress in Physics, 62, 143

\bibitem[{{Xu} \& {Lovas}(1997)}]{Xu97}
{Xu}, L.-H. \& {Lovas}, F.~L. 1997, Journal of Physical and Chemical Reference
  Data, 26, 17

\bibitem[{{Young Owl} {et~al.}(2000){Young Owl}, {Meixner}, {Wolfire},
  {Tielens}, \& {Tauber}}]{YoungOwl00}
{Young Owl}, R.~C., {Meixner}, M.~M., {Wolfire}, M., {Tielens}, A.~G.~G.~M., \&
  {Tauber}, J. 2000, \apj, 540, 886

\end{thebibliography}

\end{document}